\documentclass[aps,prstb,preprint,onecolumn,showpacs,groupedaddress]{revtex4-1}
\usepackage{graphicx}
\usepackage{epstopdf}
\usepackage{dcolumn}
\usepackage{bm}
\usepackage{amssymb}
\usepackage{amsmath}
\usepackage{mathalfa}
\usepackage[mathscr]{euscript}
\usepackage{relsize}
\usepackage{phonetic}
\begin{document}

\title{Three-dimensional analysis of the surface mode supported in $\boldmath{\check{\text{C}}}$erenkov and Smith-Purcell free-electron lasers}
\author{Yashvir Kalkal$^{1,2}$}
\email{yashvirkalkal@gmail.com}
\author{Vinit Kumar$^{1,2}$}
\email{vinit@rrcat.gov.in}

\affiliation{$^1$Homi Bhabha National Institute, Mumbai 400094, India \\
$^2$Accelerator and Beam Physics Laboratory, Raja Ramanna Centre for Advanced Technology, Indore 452013, India}

\begin{abstract}
In $\check{\text{C}}$erenkov and Smith-Purcell free-electron lasers (FELs), a resonant interaction between the electron beam and the co-propagating surface mode can produce copious amount of coherent terahertz (THz) radiation. We perform a three-dimensional (3D) analysis of the surface mode, taking the effect of attenuation into account, and set up 3D Maxwell-Lorentz equations for both these systems. Based on this analysis, we determine the requirements on the electron beam parameters, i.e., beam emittance, beam size and beam current for the successful operation of a $\check{\text{C}}$erenkov FEL.
\end{abstract}


\maketitle

%---------------------------------------------------------------------------------------------------------------------------------------------------------
\section{\label{sec:level1}Introduction}
$\check{\text{C}}$erenkov FEL (CFEL)~\cite{Danos1,Danos,Walsh3,EXPCFEL0,EXPCFEL1,EXPCFELD1,EXPCFELD2,EXPCFEL2,EXPCFEL3,Seo,Owens1,Owens2,EXPCFELD3,EXPCFELD4,Brau1,Li2,Li3,CFEL} and Smith-Purcell FEL (SP-FEL)~\cite{Levi,GratingFEL,Urata,Song,AB1,DispersionSPFEL,VinitPRE,KimPRSTB,VinitPRSTB,VinitFEL07,EXPSPFEL1,EXPSPFEL2,Donohue3DSPFEL}, which use low-energy electron beam~\cite{EXPCFEL3,Seo,Owens1,Owens2,EXPCFELD3,EXPCFELD4,Brau1,Li2,Li3,CFEL,Urata,Song,AB1,DispersionSPFEL,VinitPRE,KimPRSTB,VinitPRSTB,VinitFEL07,EXPSPFEL1,EXPSPFEL2,Donohue3DSPFEL} are seen as compact, and tunable sources of coherent THz radiation. This radiation can be utilized for a variety of applications in material science, biophysics and industrial imaging \cite{Clery,Siegel}.

In a CFEL, an electron beam skims over the surface of a thin dielectric slab placed over an ideal conductor, and resonantly interacts with the surface electromagnetic mode supported by the system, resulting in emission of coherent radiation. SP-FEL is similar to CFEL, except that the “\textquotedblleft dielectric slab placed over an ideal conductor \textquotedblright is replaced with a metallic reflection grating. Over the last few decades, several efforts have been made to realize the generation of coherent radiation from CFEL \cite{EXPCFEL0,EXPCFEL1,EXPCFELD1,EXPCFELD2,EXPCFEL2,EXPCFEL3,Owens2,EXPCFELD3,EXPCFELD4} and SP-FEL ~\cite{Urata,EXPSPFEL1,EXPSPFEL2} systems. The coherent emission of radiation with average power of the order of tens of $\mu$W was reported from a single slab based CFEL at ENEA Frascati Centre~\cite{EXPCFEL1,EXPCFEL2}. Subsequently, Fisch and Walsh~\cite{EXPCFEL3} used a low energy electron beam ($\sim$30-200 keV) to drive a sapphire based CFEL. An efficient and compact version of the device described in Ref.~\cite{EXPCFEL3}, which uses a very low energy ($\sim$ 30 keV) electron beam was tested at the Dartmouth college~\cite{Owens2} for the generation of THz radiation. The observed output power in the Dartmouth experiment~\cite{Owens2} was very low ($\sim$ picowatt), and authors in Ref.~\cite{Owens2} suggested that a good quality flat electron beam can be used to enhance the output power of the system. Recently, under the joint research of Osaka Sangyo university and Kansai university, experimental studies on a double slab based CFEL have been performed. Experimental studies on SP-FEL system have been performed at the Dartmouth college~\cite{Urata} and Vanderbilt university ~\cite{EXPSPFEL1} in USA, and  CEA/Cesta ~\cite{EXPSPFEL2} in France. The power level attained in all these experiments have been low. In order to obtain a copious coherent radiation from the $\check{\text{C}}$erenkov and Smith-Purcell FELs, an enhanced understanding of these systems is required; which includes analysing the realistic effects due to diffraction and attenuation of the surface mode supported by these systems, and then working out the requirement on  electron beam parameters, which is very critical for the performance of the system. Although for the case of SP-FEL, the requirement on electron beam parameters taking the effect of diffraction and attenuation~\cite{DispersionSPFEL} have been worked out in Refs.~\cite{KimPRSTB,VinitPRSTB,VinitFEL07}, a similar analysis has not been presented for the case of CFEL. In this paper, we perform such an analysis to work out the electron beam requirements for the case of CFEL, taking the effect due to diffraction and attenuation into account. While doing so, we also highlight important differences in the analyses of CFEL and SP-FEL.

A schematic of $\check{\text{C}}$erenkov FEL is shown in Fig.~1, where a flat electron beam is skimming at a height $h$ over a thin dielectric slab placed over an ideal conductor. The surface mode supported in this system is evanescent in the $x$-direction. In most of the earlier analyses \cite{Walsh3,Seo,Owens1,Li3,CFEL}, translational invariance is assumed along the $y$-direction, namely the horizontal direction. Due to this assumption, the surface mode supported by the structure has no variation in the $y$-direction. In this paper, we call such a surface mode as the non-localized surface mode, and the theory which describes the interaction of an electron beam with the non-localized surface mode as the two-dimensional (2D) theory. Recently, we have performed a rigorous 2D analysis \cite{CFEL} of a $\check{\text{C}}$erenkov FEL by setting up the Maxwell-Lorentz equations in both linear and non-linear regime. It was shown that for generation of terahertz radiation in a CFEL, the required beam profile is thin in the vertical direction, and wide in the horizontal direction. Hence, we used a flat electron beam with vanishing thickness in the vertical direction, and infinite width in the horizontal direction, in our analysis. The analytical expressions for the small-signal gain and the growth rate of CFEL were derived for this case. In a realistic situation, the electron beam size as well as radiation beam size will be finite. Size of the radiation beam will increase due to diffraction. This will affect the overlap of the radiation beam with the electron beam, resulting in reduction of the small-signal gain, as well as the saturated power obtained in the device. A realistic estimate of diffraction effects therefore requires a detailed three-dimensional (3D) analysis of the electromagnetic surface mode. 

One way to invoke three-dimensional effects is to solve the electromagnetic Helmholtz wave equation by considering the diffraction in the surface mode. Andrew and Brau \cite{Brau1} assumed the electron beam as a plasma dielectric, and solved the wave equation with diffraction effects to evaluate the growth rate of the CFEL. This approach works well in the linear regime. Growth rate was found to be decreasing on the accounts of the 3D effects as compared to the 2D analysis. In Ref.~\cite{Brau1}, the analysis was performed for an uniform electron beam having infinite vertical size, hence, the derived results are not very useful to obtain the electron beam parameters in the vertical direction. 

Another way to consider the 3D effects, which we follow in this paper, is to construct a localized surface mode by combining the plane waves propagating at different angles in the ($y,z$) plane, with a suitable weight factor. The surface mode constructed in this way is localized in the horizontal direction and represents a realistic situation. The technique of localization of electromagnetic modes by using superposition of plane waves is a standard technique in laser optics discussions also \cite{Lasers,Kimsynchroton,Primer}, and has been applied for the case of grating based FEL by Kim and Kumar \cite{KimPRSTB,VinitPRSTB,VinitFEL07}.

Our paper is organized as follows. In Sec. II~A, we review the 2D analysis of $\check{\text{C}}$erenkov FEL~\cite{CFEL}, and then perform the analysis for the 3D localized surface modes in Sec. II~B. In Sec. II~C, we set up the 3D coupled Maxwell-Lorentz equations for the system. Earlier analyses of $\check{\text{C}}$erenkov FEL in single slab geometry have neglected the effect of attenuation due to dielectric and conductor, which we include in the analysis presented in Sec. II. Next, in Sec. III A, we review the 2D analysis of Smith-Purcell FEL~\cite{VinitPRE,KimPRSTB}. We then set up the 3D coupled Maxwell-Lorentz equations for the system in Sec. III B and highlight the differences compared to the case of CFEL. Considering the effect of diffraction, the requirements on the quality of electron beam for the successful operation of such devices become very stringent, which is discussed in Sec. IV~A for the case of $\check{\text{C}}$erenkov FEL. We also discuss the techniques to relax these stringent requirements, and also the methods for production of electron beam of required quality. In Sec. IV~B, we take a specific example to show that with achievable beam quality, it should be possible to generate copious amount of terahertz radiation in a $\check{\text{C}}$erenkov FEL, even after including the three-dimensional effects, and effects due to attenuation. Finally, we discuss the results, and present some conclusions in Sec. V.
\begin{figure}[t]
\includegraphics[width=16.0 cm]{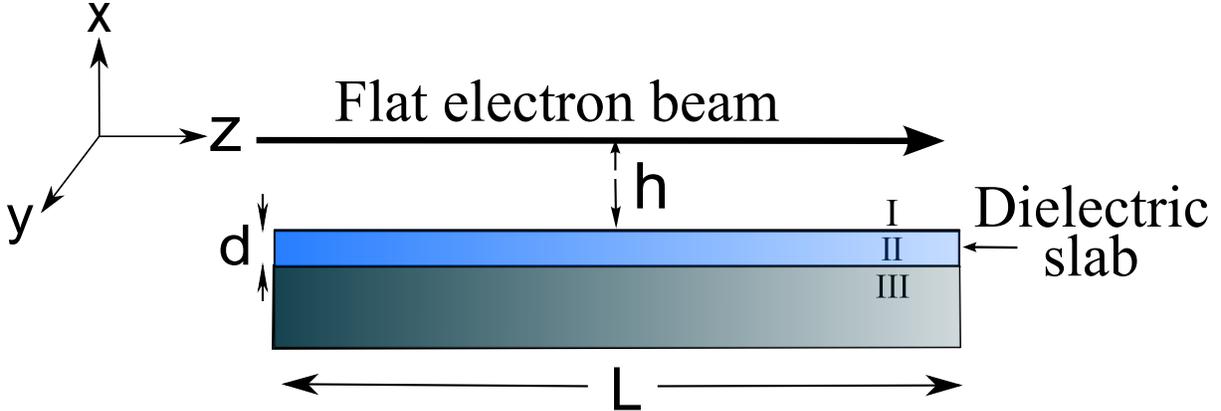}
\caption{Schematic of $\check{\text{C}}$erenkov FEL driven by a flat electron beam.}
\end{figure}

%---------------------------------------------------------------------------------------------------------------------------------------------------------
\section{SURFACE MODE ANALYSIS IN A $\check{\text{C}}$erenkov FEL}
\label{sec:2D}
In this section, we first review the 2D analysis of beam-wave interaction in a CFEL, where we have added the effect of attenuation due to finite conductivity of metallic conductor, and also the dielectric losses, which was not present in the earlier analysis~\cite{CFEL}. Details of calculation of attenuation co-efficient are given in Appendix A. Next, we set up localized surface mode, and discuss its important properties in Sec. II B. Finally, in Sec. II C, we present the derivation of three-dimensional equation for evolution of surface mode.

\subsection{Review of two-dimensional analysis}

The geometry of a $\check{\text{C}}$erenkov FEL with the coordinate system used in our analysis is shown in Fig.~1. An electron beam with vanishing thickness in the $x$-direction, is propagating with a velocity $v$ along the $z$-direction at a height $h$ above the dielectric surface. The dielectric slab has length $L$, thickness $d$ and relative dielectric constant $\epsilon$. The dielectric slab extends uniformly along the $y$-axis, and the 2D electromagnetic surface mode supported by this structure does not have any variation along the $y$-direction. To have an effective interaction with the electron beam, we want the electric field component in the $z$-direction. The appropriate surface mode for this structure can be taken as the TM mode, which has the magnetic field only in the $y$-direction, namely $H_y$ and the components of the electric field can be obtained from $H_y$ by using Maxwell equation. The longitudinal component of the electric field is given by 
\begin{eqnarray}
\label{eq:Field1}
E_z(x,z,t)=Ee^{i\psi}e^{-\Gamma x},
\end{eqnarray} 
where $E$ is the amplitude of the field at the location of the electron beam i.e., $x$=0, $\psi=k_0z-\omega t$ is the electron phase, $k_0$ is the propagation wavenumber in the $z$-direction, $\omega$=$2\pi c/\lambda$, $c$ is the speed of light, $\lambda$ is the free space wavelength and $\Gamma$=$\sqrt{k_0^2-\omega^2/c^2}$ is a positive quantity, which means that the amplitude of the surface mode decays in the $x$-direction.

The electron beam interacts with the co-propagating surface mode, and develops micro-bunching at the wavelength of the surface mode. This results in a sinusoidal component of electron beam current, having frequency $\omega$ and wavenumber $k_0$, same as that of the surface mode. This component of the beam current generates an evanescent electromagnetic wave, again having frequency $\omega$ and wavenumber $k_0$, and amplitude that decays exponentially with distance from the beam. The evanescent wave is incident on the dielectric surface, and gets reflected with a reflectivity $R=i\chi/\nu+\chi_1$, where $\chi$ and $\chi_1$ are constants, which depend on $\omega$, $k_0$, and intrinsic parameters of the system. Here, $\nu$ is the growth rate of incident evanescent wave, which is physically understood as resulting due to exponential enhancement in the micro-bunching, as the beam propagates,  due to its interaction with the surface mode. Analytical formula for the reflectivity is derived by satisfying the appropriate boundary conditions at different interfaces in Fig.~1, and this gives the expressions for $\chi$ and $\chi_1$, which are provided in Ref.~\cite{CFEL}. Note that in this approach, if $\nu=0$, the reflectivity becomes singular, for frequency $\omega$ and wavenumber $k_0$ satisfying the dispersion relation. In that case, the dielectric slab placed on the metallic conductor supports the surface electromagnetic mode in the absence of any incident wave, as expected. The reflected wave further interacts with the co-propagating electron beam, resulting in enhancement of micro-bunching, and consequently the amplitude of the incident evanescent wave further increases.  This phenomenon continues till the amplitude of the surface mode saturates due to nonlinearity.  The evolution of the amplitude $E$ of the surface mode discussed here is mathematically described by the following time-dependent differential equation, which includes attenuation of the surface mode:
\begin{eqnarray}
\label{eq:MaxCFEL}
\frac{\partial E}{\partial z}+\frac{1}{\beta_gc}\frac{\partial E}{\partial t}=\frac{-Z_0\chi}{2\beta\gamma} \frac{dI}{dy}e^{-2\Gamma h}\langle e^{-i\psi}\rangle - \alpha E,
\end{eqnarray} 
where $\beta_g=v_g/c$ is the group velocity of the surface mode in unit of $c$, which can be computed from the dispersion curve of the surface mode. The dispersion relation is here given by: $\sqrt{\epsilon \omega^2/c^2-k_0^2}\tan(d\sqrt{\epsilon \omega^2/c^2-k_0^2})$=$\epsilon \sqrt{k_0^2-\omega^2/c^2}$, which is obtained requiring the condition that the reflectivity $R$ becomes singular. The first term on the right hand side of Eq.~(\ref{eq:MaxCFEL}) represents the interaction of the electron beam with the surface mode, where $Z_0$=$1/\epsilon_0c$=377 $\Omega $ is the characteristic impedance of free space, $\epsilon_0$ is the permittivity of the free-space, $\beta = v/c$, $\gamma $ is the relativistic Lorentz factor, $I$ represents the electron beam current, $dI/dy$ is the linear current density of flat electron beam, and $\langle \cdots \rangle$ indicates averaging over the number of particles distributed over one wavelength of the evanescent surface mode. The second term on the right hand side of Eq.~(\ref{eq:MaxCFEL}) represents attenuation of the surface wave due to losses present in the dielectric and metallic structures. Note that the above equation without the attenuation term is described in Ref.~\cite{CFEL}. The attenuation coefficient $\alpha$ is sum of the dielectric attenuation coefficient $\alpha^{d}$ and ohmic attenuation coefficient $\alpha^{c}$ due to losses present in the metallic conductor. As derived in detail in Appendix A, the attenuation coefficient of the surface mode in a CFEL is given by:
\begin{eqnarray}
\label{eq:Alpha2}
\alpha=\frac{\gamma k_0Z_0\tan\delta(2-\epsilon\beta^2)+\beta\epsilon^2 k_0(1+a^2)(2R_s+\beta k_0Z_0d\tan\delta)}{2Z_0[\gamma(1+\epsilon^2a^2)+\epsilon k_0d(1+a^2)]}.
\end{eqnarray} 
Here, $\tan\delta$ represents tangent loss of the dielectric medium, $a=(\gamma/\epsilon)\sqrt{\epsilon\beta^2-1}$, $R_s=\sqrt{\mu_0\omega/2\Sigma}$ is surface resistance of the metal, $\mu_0$ is the permeability of the free space, and  $\Sigma$ represents conductivity of the metal. We would like to mention that the effect of attenuation on the performance of a single slab CFEL system has been ignored in all the previous analyses. In this paper, we will show that to obtain a meaningful gain in a CFEL system, one has to minimize the losses due to attenuation in the surface mode.

In a CFEL based on positive refractive index dielectric; the surface mode will have a positive group velocity $v_g$, and will be amplified as it propagates with the electron beam in the positive $z$-direction \cite{Brau1}. When the amplified field is fed back at the entrance of the interaction region by using an optical cavity, the system starts working as an oscillator. The electron trajectories will evolve due to interaction with the electromagnetic field of the surface mode. Here, the electron motion is assumed to be strictly in the longitudinal direction, and the equations for the evolution of energy and phase of the $i$th electron are given by \cite{CFEL}:
\begin{eqnarray}
\label{eq:energy}
\frac{\partial \gamma_i}{\partial z}+\frac{1}{\beta c}\frac{\partial \gamma_i}{\partial t}=\frac{eE}{m_0c^2}e^{i\psi_i}+\text{c.c.},~~~~
\end{eqnarray}
\begin{eqnarray}
\label{eq:phase}
\frac{\partial \psi_i}{\partial z}+\frac{1}{\beta c}\frac{\partial \psi_i}{\partial t} =\frac{\omega}{c\beta^3\gamma^2} \bigg (\frac{\gamma_i-\gamma_p}{\gamma_p}\bigg).~~~~
\end{eqnarray}
Here, $e$ is the electronic charge, $m_0$ is mass of electron, and the subscript $p$ is meant for the resonant particle, whose velocity is equal to the phase velocity of the surface mode. The behaviour of a $\check{\text{C}}$erenkov FEL is governed by the Maxwell equation given by Eq.~(\ref{eq:MaxCFEL}) together with Lorentz equations given by Eqs.~(\ref{eq:energy},\ref{eq:phase}). The set of Eqs.~(\ref{eq:MaxCFEL},\ref{eq:energy},\ref{eq:phase}) can be solved numerically to obtain the detailed behaviour of the CFEL system in both linear and non-linear regime, as described in Ref. \cite{CFEL}. In the small-signal small-gain regime, an analytical solution of Eqs.~(\ref{eq:MaxCFEL},\ref{eq:energy},\ref{eq:phase}) has been obtained for the small-signal gain as \cite{CFEL}:
\begin{eqnarray}
\label{eq:GainCFEL}
G=4\times 6.75\times 10^{-2}\times 2\pi \frac{\chi}{I_A}\frac{k_0L^3}{\beta^3\gamma^4}\frac{dI}{dy}e^{-2\Gamma h},
\end{eqnarray}
where $I_A$=$4\pi\epsilon_0m_0c^3/e$=$17.04$ kA is the Alfv$\acute{\text{e}}$n current. The factor $e^{-2\Gamma h}$ is a measure of the interaction between the electron beam and the co-propagating evanescent surface mode. Taking the effect of attenuation, there will be a round trip loss given by $(1-e^{-4\alpha L})$ in addition to the gain described by Eq.~(\ref{eq:GainCFEL}). In small-signal high-gain regime, growth rate of a CFEL using mono-energetic flat electron beam is given by \cite{CFEL}:
\begin{eqnarray}
\label{eq:GrowthCFEL}
\nu=\frac{\sqrt{3}}{2L}\bigg( 2\pi \frac{\chi}{I_A}\frac{k_0L^3}{\beta^3\gamma^4}\frac{dI}{dy}e^{-2\Gamma h} \bigg)^{1/3}.
\end{eqnarray}
Taking the effect of attenuation, the growth rate will reduce to $\nu-\alpha$.

\subsection{Localized surface mode}

Now, we consider the effect of diffraction of the surface mode in the $y$-direction, and construct the 3D localized surface mode supported in a CFEL system. As the dielectric slab is an open structure in the $y$-direction, the supported electromagnetic surface mode is expected to behave like freely propagating optical beam, and will undergo diffraction in the ($y,z$) plane. The diffracting electromagnetic surface mode can be constructed by combining the plane waves propagating at different angles in the ($y,z$) plane, with suitable weight function $A(k_y)$ in $k_y$ as:
\begin{eqnarray}
\label{eq:localized1}
E_z(x,y,z,t)=\frac{1}{\sqrt{2\pi}}\int dk_yA(k_y)e^{i(k_zz-\omega t)}e^{ik_yy}e^{-\Gamma'x}.
\end{eqnarray}
Here, $E_z$ is the longitudinal electric field, $\Gamma'$ is the attenuation constant due to evanescent nature in the $x$-direction, when the wave is propagating in the ($y,z$) plane with wavenumbers  $k_y$ and $k_z$ in $y$-direction and $z$-direction, respectively. The surface mode constructed in this way will have a variation along the $y$-direction and will represent the generalized case of surface mode given by Eq.~(\ref{eq:Field1}). The electromagnetic surface mode given by Eq.~(\ref{eq:localized1}) is mainly propagating in the $z$-direction, and undergoes diffraction in the $y$-direction.

In the CFEL system, the dielectric slab is an isotropic structure in the ($y,z$) plane. Using the property of isotropy in the $(y,z)$ plane, the dispersion relation of the surface mode propagating along the $z$-axis can be easily generalized to the case, where the surface mode is propagating along any arbitrary direction in the $(y,z)$ plane. For a given frequency $\omega$, if the phase velocity of the surface mode propagating along the $z$-axis is $v$, we obtain the following relation between $\omega$, $k_y$ and $k_z$ for a surface wave propagating in the $(y,z)$ plane:
\begin{eqnarray}
\label{eq:resonance}
\omega=v\sqrt{k_y^2+k_z^2}.
\end{eqnarray} 
The wavenumber in the longitudinal direction can be written as $k_z$=$k_0+\Delta k$, where $k_0$=$\omega/v$. By using the paraxial approximation ($k_y\ll k_z$), we obtained the following expression for $k_z$: 
\begin{eqnarray}
\label{eq:kCFEL}
k_z=k_0\bigg(1-\frac{k_y^2}{2k_0^2}\bigg).
\end{eqnarray}
Note that due to the property of isotropy in the ($y,z$) plane, $\Gamma'=\Gamma$. We can substitute Eq.~(\ref{eq:kCFEL}) for $k_z$ and  $\Gamma'=\Gamma$ in Eq.~(\ref{eq:localized1}) to obtain the localized surface mode in a CFEL as:
\begin{eqnarray}
\label{eq:localized2}
E_z(x,y,z,t)=\frac{e^{i\psi}e^{-\Gamma x}}{\sqrt{2\pi}} \int \underbrace{A(k_y)e^{-ik_y^2z/2k_0}}e^{ik_yy}dk_y.
\end{eqnarray}
Above expression for the longitudinal field appears as a Fourier transform in $k_y$ of the underbraced term. If we choose $A(k_y)$=$e^{-k_y^2/2\sigma_{k_y}^2}$, the integration in Eq.~(\ref{eq:localized2}) is Fourier transform of a Gaussian function. Gaussian functions belong to the distinct family of functions which are self-Fourier functions \cite{Selffourier}. Hence, the resultant of integration in Eq.~(\ref{eq:localized2}) is also a Gaussian function, and we obtain the intensity for the localized Gaussian mode at $x$=0 as:
\begin{eqnarray}
\label{eq:intensity}
\text{Intensity}: E_z\times E_z^* \propto e^{-\frac{y^2}{2}{\frac{2\sigma_{k_y}^2}{1+\sigma_{k_y}^4z^2/k_0^2}}}.
\end{eqnarray}
We want to emphasize that this approach can be easily generalized for higher order modes by taking Gauss-Hermite functions for $A(k_y)$, which are also self-Fourier functions, and will give higher order Gauss-Hermite modes.

Next, we analyze the transverse properties of the localized surface mode. Using Eq. (\ref{eq:intensity}), we obtain an expression for the variation of rms optical beam size $\sigma_y$ with $z$ as:
\begin{eqnarray}
\label{eq:Mode1}
 \sigma_y^2(z)=\sigma_y^2(0)\bigg(1+\frac{z^2}{Z_R^2}\bigg),
\end{eqnarray}
Here, $\sigma_y(0)$ is the rms optical beam waist at $z$=0, and $Z_R$ is the Rayleigh range which is obtained as:
\begin{eqnarray}
\label{eq:ModerangeCFEL}
Z_R=\frac{4\pi \sigma_y^2(0)}{\beta \lambda}.
\end{eqnarray}
Another quantity of interest is the product of rms beam waist size and rms angular divergence $\sigma_{\theta}$, which is given as:
\begin{eqnarray}
\label{eq:ModeconstCFEL}
\sigma_y(0)\times \sigma_{\theta}=\frac{\beta\lambda}{4\pi}.
\end{eqnarray}
Note that above expressions are similar to the standard expressions for the case of Gaussian mode propagating in free space except that $\lambda$ is replaced with $\beta \lambda$. It is well known in optics that for a Gaussian mode propagating in a uniform, isotropic medium, $\lambda$ gets replaced with $\lambda /n$ in the above formulas, where $n$ is the refractive index of the medium. Using $n = c/v_p$, where $v_p$ is phase velocity of light in the medium, $\lambda /n$ is same as $(v_p/c)\lambda$. This is thus similar to the result we obtained for the surface mode here. 

The present analysis for the localized surface mode will be used to estimate the required parameters of the electron beam for efficient working of the $\check{\text{C}}$erenkov FEL in Sec IV.

\subsection{Three-dimensional Maxwell-Lorentz equations}

Next, we will derive the three-dimensional Maxwell-Lorentz equations for the CFEL system. We start with the generalized expression for the sinusoidal component of the beam current density: $\mathcal{J}= J(x,y)e^{i(k_0z-\omega t)}\langle e^{-i\psi}\rangle+$c.c., where $J(x,y)$ is the dc current density, $\langle e^{-i\psi}\rangle$ indicates bunching of the electron beam due to interaction with the surface mode, and c.c. represents complex conjugate. We can decompose the beam current density into Fourier components as: 
\begin{eqnarray}
\label{eq:JCFEL}
\mathcal{J}=\frac{1}{\sqrt{2\pi}}\int \underbrace{\widetilde{J}(x,k_y)e^{ik_y^2z/2k_0}\langle e^{-i\psi}\rangle} e^{i\psi}e^{-ik_y^2z/2k_0}e^{ik_yy}dk_y+\text{c.c.}.~~~~
\end{eqnarray}
Note that we have cast the integral in a form such that the Fourier component of the electromagnetic field can be understood to be evolving with the Fourier component of the electron beam current density. For further calculations, we consider flat electron beam for which $J(x,y)$=$j(y)\delta(x)$ and its Fourier transform is written as $\widetilde{J}(x,k_y)$=$\widetilde{j}(k_y)\delta(x)$. The electromagnetic fields due to this current density can be obtained by using Maxwell equations under appropriate boundary conditions, and we obtain the longitudinal component of electric field as:
\begin{eqnarray}
\label{eq:field3D1}
E_z(x,y,z,t)= \frac{e^{i\psi}e^{-\Gamma x}}{\sqrt{2\pi}}\int \underbrace{A(k_y,z,t)}e^{-ik_y^2z/2k_0}e^{ik_yy}dk_y + \text{c.c.}.
\end{eqnarray}
The amplitude $A(k_y,z,t)$ of the surface mode will evolve due to interaction with the co-propagating electron beam, and we have assumed it to be a slowly varying function of $z$ and $t$. The beam-wave interaction in a CFEL system for the 2D case is discussed earlier in Sec. II A. Now, by following the same approach, and realizing that the underbraced term in Eq.~(\ref{eq:field3D1}), which is the amplitude of Fourier component of the electromagnetic field, is evolving due to interaction with the amplitude of corresponding Fourier component of the current density, denoted by the underbraced terms in Eq.~(\ref{eq:JCFEL}), we obtain following time dependent differential equation for evolution of $A(k_y,z,t)$:
\begin{eqnarray}
\label{eq:MaxCFEL3D1}
\frac{\partial A}{\partial z}+\frac{1}{\beta_gc}\frac{\partial A}{\partial t}=
\frac{-Z_0\chi}{2\beta\gamma} \widetilde{j}(k_y)e^{ik_y^2z/2k_0} e^{-2\Gamma h}\langle e^{-i\psi}\rangle-\alpha A.~~~~
\end{eqnarray}
By taking Fourier transform with respect to $k_y$ in Eq.~(\ref{eq:MaxCFEL3D1}), and using the fact that $Ae^{-ik_y^2z/2k_0}$ is the Fourier transform of the longitudinal surface field; we obtain the following differential equation for the 3D surface mode:
\begin{eqnarray}
\label{eq:MaxCFEL3D2}
~\frac{\partial{E}}{\partial{z}}-\frac{i}{2k_0}\frac{\partial^2{E}}{\partial y^2}+\frac{1}{\beta_gc}\frac{\partial{E}}{\partial{t}}=\frac{-Z_0\chi}{2\beta\gamma}\frac{dI}{dy}e^{-2\Gamma h}\langle e^{-i\psi}\rangle-\alpha E.~~~~
\end{eqnarray}
Here, $E$ is the amplitude of longitudinal field $E_z$, and $dI/dy$ is the linear current density of the flat beam. The second term on the left hand side of above equation represents diffraction of the surface mode and allows us to study the transverse profile of the optical beam. In an approximate way, the effect of partial overlap between the electron beam and optical mode can be considered in the numerical solutions of 2D Maxwell-Lorentz equations by writing the linear current density $dI/dy$ as $I/\Delta y$, where $\Delta y$ is the electron beam width, and is replaced with the effective optical beam width determined using Eq.~(\ref{eq:ModerangeCFEL}).

%---------------------------------------------------------------------------------------------------------------------------------------------------------
\section{SURFACE MODE ANALYSIS IN A SMITH-PURCELL FEL}
The essential features of the analysis of surface mode supported in a SP-FEL system have been worked out earlier in Refs.~\cite{VinitPRE,KimPRSTB,VinitPRSTB,VinitFEL07}. Here, we further elaborate to highlight interesting difference in the analyses of CFEL and SP-FEL systems. We first review the 2D analysis, which is followed by the 3D analysis of the surface mode, where we provide derivations of some important results and present a comparison with the case of CFELs. 
\subsection{Review of two-dimensional analysis}
As shown in Fig.~2, a metallic reflection grating is used as the  slow wave structure in an SP-FEL. Here, the supported surface mode is a combination of Floquet space harmonics, since the grating is a periodic structure. The zeroth-order component will show the strongest interaction with the electron beam, and has similar structure as given by Eq.~(\ref{eq:Field1})~\cite{VinitPRE}. Similar to the case of CFEL described in the previous section, the electron beam interacts with the co-propagating surface mode, and the dynamical equation for the evolution of the amplitude of surface mode is given by~\cite{VinitPRE,KimPRSTB}:
\begin{eqnarray}
\label{eq:MaxSPFEL}
\frac{\partial E}{\partial z}-\frac{1}{\beta_gc}\frac{\partial E}{\partial t}=\frac{Z_0\chi}{2\beta\gamma}\frac{dI}{dy}e^{-2\Gamma h}\langle e^{-i\psi}\rangle + \alpha E.
\end{eqnarray}
Here, the calculation of $\chi$ requires numerical evaluation of $R$ as a function of growth rate for a given value of ($\omega$, $k_0$) of the surface mode, and details regarding the procedure for this calculation are described in Ref.~\cite{VinitPRE}. Evaluation of attenuation coefficient $\alpha$ requires calculation of heat dissipation at metallic surfaces for the given surface mode, which has been discussed for the case of SP-FEL in Ref.~\cite{DispersionSPFEL}. Note the difference in sign of terms containing $\beta_g$,  $\chi$ and $\alpha$ in above equation as compared to Eq.~(\ref{eq:MaxCFEL}) for CFEL. This is due to the fact that SP-FEL has negative group velocity for the surface mode, as described in Ref.~\cite{VinitPRE}, whereas the group velocity is positive in case of the CFEL system. The negative group velocity for SP-FEL makes it a backward wave oscillator, and oscillations build up when the linear current density $dI/dy$ exceeds a threshold value $dI_s/dy$~\cite{VinitPRE,KimPRSTB}:
\begin{eqnarray}
\label{eq:StartSPFEL}
\frac{dI}{dy}>\frac{dI_s}{dy}=\mathcal{J}(\eta)\frac{I_A\beta^4\gamma^4}{2\pi \chi kL^3}e^{2\Gamma h}.
\end{eqnarray}
Here, $\mathcal{J}(\eta)$ represents dimensionless start current as a function of the loss parameter $\eta=\alpha L$, and $k$=$\omega/c$, and is evaluated in Ref.~\cite{VinitFEL06}.

\begin{figure}[t]
\includegraphics[width=16.0 cm]{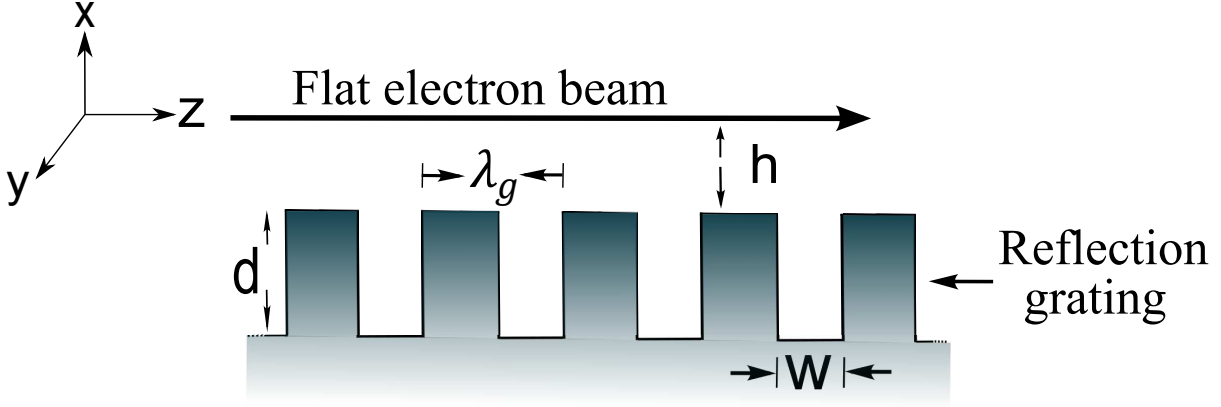}
\caption{Schematic of a Smith-Purcell FEL, using a flat electron beam.}
\end{figure}

After having briefly discussed the 2D analysis, we next discuss the localized surface mode, and set up the 3D coupled Maxwell-Lorentz equations for the SP-FEL system.

\subsection{Localized surface mode and three-dimensional Maxwell-Lorentz equations}
In order to construct the localized surface mode, we need to combine the plane waves propagating at different angles in the ($y,z$) plane with suitable weight function. In order to perform this calculation, we need to know the 3D dispersion relation i.e., the dependence of $\omega$ on $k_z$, for different values of $k_y$. As discussed in Sec. II A, the dispersion relation is obtained by deriving the condition for singularity in $R$. In an SP-FEL system, the calculation of $R$ of the grating for growing evanescent waves for the 2D case was carried out by Kumar and Kim~\cite{VinitPAC05}, and dispersion relation of the surface mode was obtained. This analysis was then extended by including exp$(ik_yy)$ type variation in the EM field, and reflectivity was evaluated ~\cite{VinitFEL07} for the 3D case. A remarkable observation in their analysis is that if we replace $\omega$ by $\sqrt{\omega^2-c^2k_y^2}$ in the expression for the reflectivity for the case $k_y$=0; we obtain the reflectivity for the 3D case, where a finite value of $k_y$ is considered \cite{VinitFEL07,Donohue3DSPFEL}. This is because here we have electromagnetic field present only in one medium, i.e., the space above the surface of the reflection grating (including the grooves of the grating), which is in vacuum.  Due to this feature, the expression for reflectivity has terms like ($\omega^2-c^2k_z^2$) in the 2D case, which  can be simply replaced with ($\omega^2-c^2k_y^2-c^2k_z^2$) for the 3D case. This amounts to replacing $\omega$ in the 2D case by $\sqrt{\omega^2-c^2k_y^2}$, to determine the dispersion relation for 3D case with finite $k_y$.

It is important here to note the difference between the dispersion relation of SP-FEL and CFEL system. A careful observation of the 2D dispersion relation of CFEL: $\sqrt{\epsilon \omega^2/c^2-k_0^2}\tan(d\sqrt{\epsilon \omega^2/c^2-k_0^2})$=$\epsilon \sqrt{k_0^2-\omega^2/c^2}$~\cite{CFEL} indicates that the simple \textquotedblleft replacement rule\textquotedblright~as in the case of SP-FEL system i.e., replacing $\omega$ in 2D dispersion relation with $\sqrt{\omega^2-c^2k_y^2}$ will not give us the 3D dispersion relation. This is because here, the electromagnetic field in CFEL is present in vacuum as well as in the dielectric medium, unlike the SP-FEL case; and therefore terms like ($\omega^2-c^2k_o^2$) as well as ($\omega^2-c^2k_o^2/\epsilon$) appear in the 2D dispersion relation. In the case of CFEL, the \textquotedblleft isotropic nature\textquotedblright~of the dielectric slab in $(y,z)$ plane facilitates us to analyze the diffraction in the surface mode as described in the previous section. The grating structure used in the SP-FEL system has grooves along the surface in the transverse direction, and lacks isotropic behaviour in the $(y,z)$ plane. Due to this difference, the optical properties of the surface mode in SP-FEL system are different as compared to CFEL system, as elaborated below.

We now discuss the construction of localized surface mode in SP-FEL. Due to the replacement rule: $\omega_{3D}(k_y)=\sqrt{\omega_{2D}^2+(ck_y)^2}$, we write the longitudinal wavenumber $k_z$ as:
\begin{eqnarray}
\label{eq:Kz1}
k_z=k_0+\frac{\partial k}{\partial \omega}\bigg\vert_{k_y=0}\Delta \omega,
\end{eqnarray}
where $\Delta \omega$=$\omega_{2D}-\omega_{3D}$ and the term $\partial k/\partial \omega$ at $k_y$=0 is identified as $(-1/\beta_gc)$. Using these results along with the paraxial approximation in Eq.~(\ref{eq:Kz1}), we obtain:
\begin{eqnarray}
k_z=k_0\bigg(1+\frac{k_y^2}{2\beta\beta_gk_0^2}\bigg).
\end{eqnarray}
Here, we note the difference between above equation and corresponding equation [Eq.~(\ref{eq:kCFEL})] for the case of CFEL. On the account of 3D effects, the magnitude of change in the longitudinal wavenumber ($\vert k_z-k_0\vert$) is given by $\beta\lambda k_y^2/4\pi$ for the CFEL, and $\lambda k_y^2/4\pi\beta_g$ for the SP-FEL case. It can be seen that in this term, $\beta\lambda$ in case of CFEL is replaced with $\lambda/\beta_g$ for the case of SP-FEL.

Next, by satisfying the wave equation for the electromagnetic field, we obtain the expression for $\Gamma'$ as:
\begin{eqnarray}
\Gamma'=\Gamma\bigg(1+\frac{k_y^2(1+\beta\beta_g)}{2\beta\beta_g\Gamma^2}\bigg).
\end{eqnarray}
By following an approach similar to the one described in the Sec II B; the analysis for the localized surface mode supported by the grating structure is performed. The Rayleigh range for the optical surface mode is obtained as~\cite{KimPRSTB}:
\begin{eqnarray}
\label{eq:ModerangeSP}
Z_R=\frac{4\pi\beta_g \sigma_y^2(0)}{\lambda},
\end{eqnarray}
where $\sigma_y(0)$ is the rms beam size at the waist. Under paraxial approximation, the product of rms beam waist size and rms divergence is given by~\cite{KimPRSTB}:
\begin{eqnarray}
\label{eq:ModeconstSP}
\sigma_y(0)\times \sigma_{\theta}=\frac{\lambda}{4\pi\beta_g}.
\end{eqnarray} 
Note that the Eqs.~(\ref{eq:ModerangeSP}-\ref{eq:ModeconstSP}) have dependence on the group velocity, while equivalent quantities in the CFEL system [Eqs.~(\ref{eq:ModerangeCFEL}-\ref{eq:ModeconstCFEL})] have dependence on the phase velocity of the surface mode. In these expressions, the term $\beta \lambda$ in the case of CFEL is replaced with $\lambda/\beta_g$ in the case of SP-FEL, as expected. We emphasize that this difference arises due to fundamental difference in the way the dispersion relation for the two systems  gets modified for the 3D case, which we have explained. Due to this nature, it can be seen that the diffraction effects are more prominent in case of SP-FEL as compared to the CFEL. The length $L$ of the grating in case of SP-FEL has to be kept small to maintain sufficient interaction of the surface mode with the co-propagating electron beam. 

Next, the expression for $k_z$ and $\Gamma'$ can be used in Eq.~(\ref{eq:localized1}) to set up the three-dimensional electromagnetic surface mode for SP-FEL. By following the procedure described in the Sec. II C, the following time dependent three-dimensional differential equation for the evolution of the surface mode in a SP-FEL is obtained \cite{VinitFEL07}:
\begin{eqnarray}
\label{eq:MaxSP3D}
\frac{\partial E}{\partial z}+\frac{i}{2\beta\beta_gk_0}\frac{\partial^2 E}{\partial y^2}-\frac{1}{\beta_gc}\frac{\partial E}{\partial t}=\frac{Z_0\chi}{2\beta\gamma}\frac{dI}{dy}e^{-2\Gamma h}\langle e^{-i\psi}\rangle+\alpha E.~~~~
\end{eqnarray}
Note the difference in the second term of above equation as compared to the corresponding term in Eq.~(\ref{eq:MaxCFEL3D2}) for the CFEL. Here, a factor $\beta\beta_g$ is appearing, which shows large diffraction in the surface mode in an SP-FEL, as compared to the CFEL. Note that although the diffraction term in the above equation has the same form as in the case of undulator based FEL~\cite{Braubook}, the free space wavelength $\lambda$ appearing in the this term for undulator based FEL is replaced with $\beta \lambda$ in CFEL, and $\lambda/\beta_g$ in SP-FEL, and this is an important finding of our analysis. We would like to mention here that although the Ref. \cite{VinitFEL07} describes Eq.~(\ref{eq:MaxSP3D}), it does not elaborate on the procedure to derive this equation, which is provided in this paper.
%-----------------------------------------------------------------------------------------------------------------------------------------------------
\section{Electron beam requirements and its production for CFEL}
\label{sec:electronbeam}
In this section, based of the analysis of surface mode presented in Sec. II B, we will work out the electron beam requirements for successful operation of a CFEL. Similar analysis for SP-FEL has already been worked out in Refs.~\cite{KimPRSTB,VinitPRSTB}, which we extend to the case of CFEL in this section.

\subsection{Theoretical analysis}
The electron beam distribution in the four dimensional phase space ($x,\psi,y,\phi$) is assumed to be Kapchinskij-Vladimirskij (KV) distribution \cite{Chaobook}, where $x$ and $y$ are the vertical and horizontal coordinates, respectively and $\psi$ and $\phi$ represents vertical and horizontal angles, respectively. The electron beam distribution is assumed to have half widths ($\Delta x, \Delta\psi, \Delta y, \Delta\phi$) at the middle of the dielectric slab, and the half widths are two times the rms values $(\sigma_x,\sigma_{\psi},\sigma_y,\sigma_{\phi})$. Thus, $\Delta x=2\sigma_x$, $\Delta\psi=2\sigma_{\psi}$, $\Delta y=2\sigma_y$, and $\Delta\phi=2\sigma_{\phi}$. The geometric rms emittance in the $y$-direction is therefore given by $\bm{\varepsilon}_y^0$=$(1/4)\Delta y\Delta \phi$. The  Courant-Snyder envelope $\beta_y$, also known as the beta function in the $y$-direction, is defined as $\beta_y$=$\sigma_y^2/\bm{\varepsilon}_y^0$. Similar quantities are defined with the subscript $x$ in the $x$-direction.

Let us first look for the requirements on the electron beam in the $y$-direction. The product of rms beam size $\sigma_y(o)$ and divergence $\sigma_{\theta}$ for the surface mode supported in CFEL is given by $\beta \lambda/4\pi$. Now to ensure that electron beam envelope is within the envelope of optical beam, the rms unnormalized emittance is required to be less than this product. Applying this for the case of CFEL, we get:
\begin{eqnarray}
\label{eq:EyCFEL}
\bm{\varepsilon}_y \leq \frac{\beta^2\gamma\lambda}{4\pi},
\end{eqnarray}
where $\bm{\varepsilon}_y=\beta\gamma\bm{\varepsilon}_y^0$ is the normalized beam emittance in the $y$-direction. Next, the half width $\Delta y$ of the electron beam, which is taken same as the half width $2\sigma_y$ of the optical beam, is chosen by requiring that the Rayleigh range $Z_R$ is equal to the interaction length $L$. This choice of $Z_R$ ensures that the variation in the rms optical beam size over the interaction length is within 10 \%, as can be seen by putting $z=L/2$ ($z=0$ corresponds to middle of the dielectric slab, and $z=\pm L/2$ corresponds to the end points), and $Z_R = L$ in Eq.(\ref{eq:Mode1}). Now, by using Eq.~(\ref{eq:ModerangeCFEL}), we find $\sigma_y=\sqrt{\beta\lambda Z_R/4\pi}$, and by inserting it in above mentioned condition; we obtain:
\begin{eqnarray}
\label{eq:DyCFEL}
\Delta y=\sqrt{\frac{\beta\lambda L}{\pi}}.
\end{eqnarray}

Let us now discuss the required electron beam parameters in the $x$-direction. In the view of the exponential factor $e^{-2\Gamma h}$ in Eq.~(\ref{eq:GainCFEL}), it is desirable that the height $h$ of the electrons should satisfy $h\leq 1/2\Gamma$ for the sufficient interaction between the electron beam and the co-propagating surface mode. Here, $\Gamma=2\pi/\beta\gamma\lambda$. Assuming that the electron beam is propagating over the dielectric slab such that its centroid is at height $h$ and its lower edge just touches the dielectric surface, we can take the half width $\Delta x$ of the electron beam same as $h=1/2\Gamma$, and obtain:
\begin{eqnarray}
\label{eq:DxCFEL}
\Delta x=\frac{\beta\gamma\lambda}{4\pi}.
\end{eqnarray}        
This implies that rms electron beam size $\sigma_x = \Delta x/2 = \beta\gamma\lambda/8\pi$ at the middle of the dielectric slab. In order to ensure that the variation in $\sigma_x$ over the interaction length is less than $\sim 10 \%$, we require $\beta_x \ge L$. Using these two conditions and the relation that $\bm{\varepsilon}_x^0 = \sigma_x^2/\beta_x$,  we obtain
\begin{eqnarray}
\label{eq:ExCFEL}
\bm{\varepsilon}_x\leq\frac{\beta^3\gamma^3\lambda^2}{64\pi^2L},
\end{eqnarray}
where $\bm{\varepsilon}_x=\beta\gamma\bm{\varepsilon}_x^0$ is the normalized beam emittance in the $x$-direction. The condition on the normalized beam emittance $\bm{\varepsilon}_x$ comes out to be very stringent. As discussed in detail in the next sub-section, a flat electron beam with transverse emittance ratio, $\bm{\varepsilon}_y/\bm{\varepsilon}_x\simeq 1000$ is required for the operation of a practical CFEL. This value is roughly 10 times higher than the value achieved in a recent experiment \cite{FlatphaseE1}.

The stringent requirement on the emittance of a flat electron beam can be relaxed by introducing an external focusing by either using a wiggler field~\cite{Flatfocusing1,Flatfocusing2} or by using a solenoid field~\cite{LiSolenoid,DonohueSolenoid}. Details of the two schemes are described in Ref.~\cite{VinitPRSTB} for the case of SP-FEL. Both the schemes are applicable for the case of CFEL also. We briefly discuss these schemes for CFEL case, and give the relevant mathematical formulas here.
\subsubsection{Focusing of flat beam by using wiggler field}
In the first scheme, where a wiggler magnetic field is used for external focusing; the flat electron beam is generated by a novel phase space technique \cite{Flatproduction,KimPRSTB}. In this technique, a cathode is placed in an axial magnetic field to produce a round beam, and then the angular momentum of beam is removed by using a set of quadrupoles. Finally, a flat electron beam is obtained with transverse emittance ratio~\cite{KimPRSTB}:
\begin{eqnarray}
\label{eq:Eratio}
\frac{\bm{\varepsilon}_y}{\bm{\varepsilon}_x}=\bigg(\frac{eB}{m_0c}\frac{r_t^2}{4\bm{\varepsilon}_I}\bigg)^2,
\end{eqnarray}  
where $B$ represents the magnetic field at the cathode, $r_t$ is the radius of the thermionic cathode, and $\bm{\varepsilon}_I$=$\sqrt{\bm{\varepsilon}_x\bm{\varepsilon}_y}$ is the initial beam emittance of the round beam. The radius $r_t$ is related to the initial emittance as $r_t$=$2\bm{\varepsilon}_I/\sqrt{k_BT/m_0c^2}$ \cite{KimPRSTB}, where $k_B$ is Boltzmann's constant and $T$ is the absolute temperature of the thermionic cathode. The magnetic field required to produce an electron beam with the desired transverse emittance ratio is evaluated by using Eq.~(\ref{eq:Eratio}) as $B$=$k_BT/e\bm{\varepsilon}_x c$. Note that $B$ is independent of $\bm{\varepsilon}_y$. The current density $J_t$ at the cathode for a given beam current $I$, is $J_t$=$I/\pi r_t^2$~\cite{KimPRSTB}.
\begin{figure}[t]
\includegraphics[width=16.0 cm]{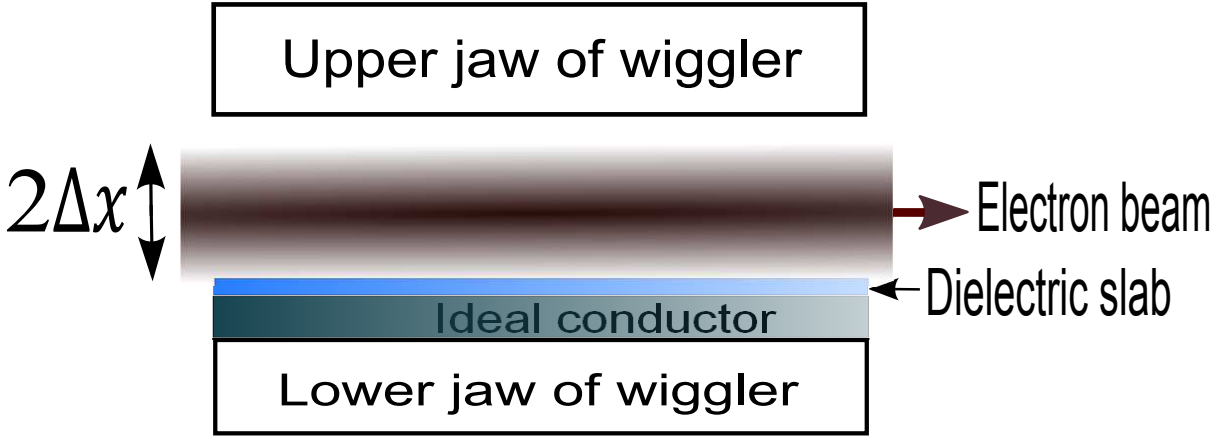}
\caption{Schematic of external focusing in a $\check{\text{C}}$erenkov FEL using a wiggler.}
\end{figure}

In Fig.~3, we have shown the schematic for focussing of the above mentioned flat beam in a CFEL by using a wiggler with a parabolic pole shape. In the presence of wiggler magnetic field, the electron beam will be focused in both $x$ and $y$ directions. By neglecting the space charge effect in envelope equation, the matched rms beam size in the $x$ and $y$ directions are obtained as~\cite{VinitPRSTB}:
\begin{eqnarray}
\label{eq:USigmaCFEL}
\sigma_{x,y}=2^{1/4}\sqrt{\frac{\bm{\varepsilon}_{x,y}}{a_uk_{x,y}}}.
\end{eqnarray}
Here, $a_u=eB_u/k_u m_0c$, $B_u$ represents the peak value of the magnetic field in the $x$-direction, along the $z$-axis, $k_u$=$2\pi/\lambda_u$, $\lambda_u$ is the wiggler period, and $k_x$ and $k_y$ represent spatial frequency of the wiggler field in $x$ and $y$-direction, respectively. We require $k_u^2$ = $k_x^2+k_y^2$ to satisfy the Maxwell equations. In Eq.~(\ref{eq:USigmaCFEL}), we choose $\sigma_y$=$1/2\Delta y$ and $\sigma_x$=$1/2\Delta x$, where $\Delta y$ and $\Delta x$ are given by Eqs.~(\ref{eq:DyCFEL}) and (\ref{eq:DxCFEL}), respectively; and find the appropriate value of $a_u,k_x$ and $k_y$ for a given values of emittances ($\bm{\varepsilon}_x,\bm{\varepsilon}_y$), such that the beam sizes are matched inside a wiggler, and thus maintain a constant size throughout the wiggler. For a typical set of parameters of a CFEL, the focusing requirement in the vertical direction is very strong as compared to the horizontal direction; we can therefore choose $k_y$=0 and $k_x$=$k_u$. It is clear from Eq.~(\ref{eq:USigmaCFEL}) that for a matched beam size, one can tolerate relaxed vertical emittance by choosing the higher value of the peak undulator magnetic field $B_u$.  Note that in case of external focusing in the vertical direction, we do not require to satisfy Eq.~(\ref{eq:ExCFEL}).
\subsubsection{Focusing of flat beam by using solenoid field}
In the second scheme, a solenoid magnetic field is used to focus a low energy flat electron beam. The required flat beam is generated using an elliptically shaped, planar thermionic cathode with major axis $\Delta y_c$ = $\Delta y$, and minor axis $\Delta x_c$ = $\Delta x$. The normalized thermal emittances for the thermionic cathode are given by $(\bm{\varepsilon}_x,\bm{\varepsilon}_y)$ = $0.5(\Delta x_c, \Delta y_c)\sqrt{k_BT/m_0c^2}$, and current density at cathode, corresponding to current $I$ is given by $J_c=I/\pi\Delta x_c\Delta y_c$~\cite{KimPRSTB}. For generating a flat beam, the vertical dimension of the cathode is very small compared to the horizontal dimension, and such a cathode is called a line cathode. The dielectric slab together with the line cathode is placed inside the solenoid, and in order to ensure that the beam does not rotate, the electron beam is generated in the uniform field region of the solenoid, and it remains inside the solenoid, while propagating over the surface of the dielectric slab. The solenoid field strength required to focus a flat beam can be evaluated with the condition that the Larmor radius should be much smaller than the verticle rms beam size $\sigma_x$ \cite{VinitPRSTB}, which gives us the following expression for the axial magnetic field $B(0)$ near the cathode ~\cite{VinitPRSTB}:
\begin{eqnarray}
\label{eq:Solenoidfield}
B(0)\gg \frac{m_0c\bm{\varepsilon}_x}{e\sigma_x^2}.
\end{eqnarray} 
The nonuniformity in the longitudinal on-axis magnetic field gives rise to rotation $\theta$ to the flat beam given by~\cite{VinitPRSTB}: 
\begin{eqnarray}
\label{eq:Rotationangle}
\theta(z)=\frac{z\omega_L}{3\beta c}\frac{\Delta B(z)}{B(0)},
\end{eqnarray} 
where $\Delta B(z)$=$B(z)-B(0)$, and $\omega_L$ is the Larmor frequency. Here, it is assumed that the cathode is placed at the centre of the solenoid ($z=0$), where the field is maximum, and the variation of the quantities in the radial direction is assumed to be very slow. We have to ensure that the electron beam does not rotate significantly such that the flat beam nature is preserved.

Clearly, both the external focusing techniques allow us to tolerate larger emittance of the electron beam, but these also increase the deleterious effects of the velocity spread. Due to external focussing, the spread in the longitudinal velocity is given by~\cite{VinitPRSTB}:
\begin{eqnarray}
\label{eq:velocityspread}
\frac{\Delta\beta}{\beta}\sim\frac{\bm{\varepsilon}_x^2}{2\beta^2\gamma^2\sigma_x^2}.
\end{eqnarray} 
Focusing in the $y$-direction will also give similar contribution to the velocity spread. Spread in the longitudinal velocity is equivalent to the spread in the energy. The maximum energy spread that can be tolerated in a CFEL corresponds to the phase mismatch of $\pi$ between the electrons and the co-propagating surface mode at the exit of the interaction region, or equivalently $\Delta\beta L/\beta$=$\beta\lambda/2$.  This condition gives us the maximum value of emittance which can be tolerated by the system as:
\begin{eqnarray}
\label{eq:Extolerance}
\bm{\varepsilon}_x < \sigma_x\sqrt{\frac{\beta^3\gamma^2\lambda}{L}}.
\end{eqnarray}
With the external focussing, we can increase the length $L$ of the dielectric slab to obtain the maximum gain. However, increment in $L$ will restrict the maximum emittance that can be tolerated; a condition given in Eq.~(\ref{eq:Extolerance}).~We need to choose an optimum length of the system for which the deleterious effect due to resulting energy spread are significantly less.

Finally, we summarize the procedure for the optimization of the focusing strength and the emittance as follows: We first choose the vertical beam size from Eq.~(\ref{eq:DxCFEL}) for the given parameters of a CFEL, and then we choose the maximum focussing strength by using Eq.~(\ref{eq:USigmaCFEL}) (in case of wiggler focusing) or using Eq.~(\ref{eq:Solenoidfield}) (in case of solenoid focussing) to attain the maximum tolerance on the vertical emittance; keeping in mind that the constraint is given by Eq.~(\ref{eq:Extolerance}).

\subsection{An example case}
For an example case of a practical CFEL, we take parameters of Dartmouth experiment~\cite{Owens2}, and optimize them in accordance with the analysis given in the earlier sections. In the Dartmouth experiment~\cite{Owens2}, an electron beam with 1 mA current, and with energy range 30-40 keV, was allowed to pass over the dielectric slab of thickness 350 $\mu$m. Two different materials, GaAs ($\epsilon$=$13.1$) and sapphire ($\epsilon$=$9.6$) were used for the dielectric slab, and a silver polished copper metal was used to support the dielectric slabs. We choose sapphire as dielectric material, which has very low tangent loss ($\tan\delta\le 10^{-4} $) as compared to GaAs ($\tan\delta \simeq 10^{-3}$)~\cite{dielectriclossdata}. We take electron beam energy as 46.5 keV, corresponding to $\beta$=$0.4$ ($\beta\gamma$=0.44), which is well above the threshold condition ($\beta_t$=$1/\sqrt{\epsilon}$=0.32) for the generation of $\check{\text{C}}$erenkov radiation in sapphire. For these parameters, we find $\lambda$=2.7 mm, $\beta_g$= 0.27 and $\chi$=317 per m by using the analysis given in Ref.~\cite{CFEL}. The conductivity of silver metal at 300 K is given by $6.3\times 10^7/\Omega$-m~\cite{Silverresitivity}, for which the attenuation coefficient, $\alpha=2.2$ per m, as calculated by using Eq.~(\ref{eq:Alpha2}). Note that the dielectric losses in sapphire are negligible for the chosen parameters. In the context of three-dimensional analysis, the linear current density $dI/dy$, which is needed to evaluate the gain and the growth rate of the system by using Eqs.~(\ref{eq:GainCFEL}) and (\ref{eq:GrowthCFEL}), respectively, can be interpreted as the peak value at the middle of the electron beam distribution, which is given by~\cite{KimPRSTB}
\begin{eqnarray}
\label{eq:linear}
\frac{dI}{dy}=\frac{I}{\pi\Delta y/2},
\end{eqnarray}
for KV distribution discussed in Sec. IV A. The effective electron beam width in the $y$-direction is thus taken as $\pi$ times the rms optical beam waist size $\sigma_y(0)$, and can be evaluated by using Eq.~(\ref{eq:DyCFEL}). The electron beam height, $h$ is taken as half beam width $\Delta x$ in the $x$-direction. The length of dielectric slab was taken as 1 cm in the Dartmouth experiment. For $L$=1 cm, we find small signal gain around 0.03 $\%$, which is too low to overcome losses present in the system, as the round trip loss ($1-e^{-4\alpha L}$) is around 8.4 $\%$. To get an appreciable gain, we take $L$=5 cm, and increase the electron beam current from 1 mA to 35 mA. By using the modified parameters in Eqs.~(\ref{eq:GainCFEL}) and (\ref{eq:GrowthCFEL}), we find gain as 50 $\%$ and grwoth rate as $21.2$ per m, respectively. With the increased length, power loss due to attenuation also increases. To reduce the attenuation, we propose that the silver metal, which supports the dielectric slab, should be kept at very low temperature i.e., at 77 K, which is the boiling point temperature of liquid nitrogen. The conductivity of silver at 77 K is about $3.3\times 10^8/\Omega$-m~\cite{Silverresitivity}, which gives us $\alpha$=0.97 per m, and round trip loss of 17.6 $\%$ over an interaction length of 5 cm. In order to attain saturation, the system is operated in the oscillator configuration, where a set of mirrors are used to provide an external feedback. One mirror is assumed to be ideal with 100 $\%$ reflectivity of the field amplitude, and other has reflectivity of 98 $\%$. The Maxwell-Lorentz equations have been solved numerically by using the Leapfrog scheme~\cite{CFEL} to obtain the power in the surface mode. The power builds up slowly and saturates after 100 number of passes, as shown by solid curve in Fig~4. At saturation, we obtain the output power as 7.2 W. The input electron beam power is 1.6 kW, which gives us efficiency of 0.44 $\%$. Figure 4 also shows the output power (dashed curve) of a CFEL, where ohmic losses are assumed to be zero. In this case, the CFEL system gives 41.3 W output power on saturation, with an efficiency of 2.5 $\%$. Note that the presence of ohmic losses on the metallic surface severely affects the output power and efficiency of a CFEL system, and one has to optimize the system for minimum losses.

The requirements on electron beam sizes are evaluated by using Eqs.~(\ref{eq:DyCFEL}) and (\ref{eq:DxCFEL}), as $\Delta y$=4.2 mm in the $y$-direction and $\Delta x$= 94 $\mu$m in the $x$-direction, respectively. By using Eq.~({\ref{eq:ExCFEL}}), we find that an electron beam with normalized vertical emittance $\bm{\varepsilon}_x \leq 1.9\times 10^{-8}$ m-rad is needed in the absence of any external focussing, which is a very stringent requirement. In the horizontal direction, the condition on beam emittance is quite relaxed as an electron beam with $\bm{\varepsilon}_y\leq 3.8\times 10^{-5}$ m-rad is required, which is calculated from Eq.~({\ref{eq:EyCFEL}}). If we take $\bm{\varepsilon}_y = 1.9\times 10^{-5}$ m-rad, which is two times less than the maximum allowed value, then a flat electron beam with transverse emittance ratio $\bm{\varepsilon}_y/\bm{\varepsilon}_x$=1000 is required for the operation of the CFEL. In Dartmouth experiment \cite{Owens2}, these conditions have been clearly violated, where a round electron beam having large vertical emittance was used to drive the $\check{\text{C}}$erenkov FEL. As discussed in the previous sub-section, the transverse emittance ratio of 100 has been achieved experimentally \cite{FlatphaseE1}, and it is feasible to extend this ratio to 400 \cite{Flatphase3}

\begin{figure}[t]
\includegraphics[width=16.0 cm]{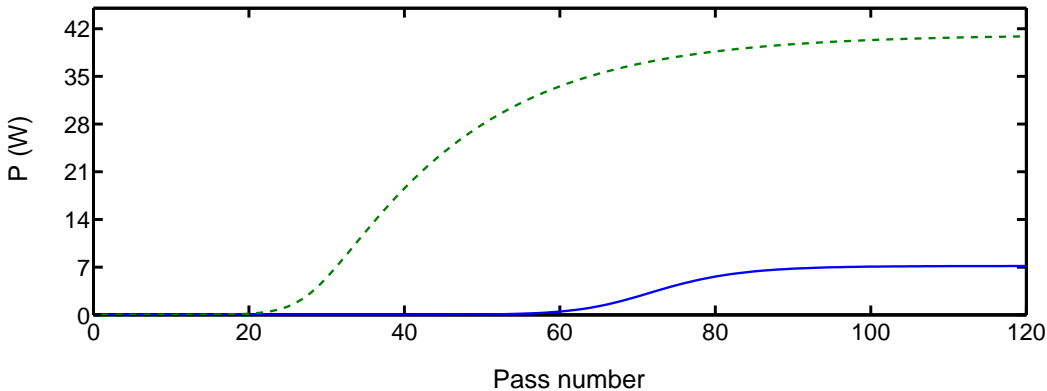}
\caption{Plot of output power as a function of pass number for the optimized parameters of a $\check{\text{C}}$erenkov FEL discussed in the text. The dashed curve represents the case, where ohmic losses are assumed to be zero, and solid curve shows the output power with finite ohmic losses in the system at 77 K temperature. The linear current density ($dI/dy$) of the electron beam is taken as 5.4 A/m.}
\end{figure}

To relax the stringent requirement on the electron beam emittance, external focusing can be provided by using a wiggler field as described in Sec.~IV~A\textit{1}. Here, we will take an explicit example to perform the calculations for the analytical results discussed in Sec.~IV~A\textit{1}. We assume a round electron beam with initial normalized emittance $\bm{\varepsilon}_I$=$1\times{10^{-6}}$ m-rad, which is easily achievable. The flat electron beam can be produced by using round to flat beam transformation as discussed earlier, and under this transformation $\bm{\varepsilon}_I$=$\sqrt{\bm{\bm{\varepsilon}_x\varepsilon}_y}$. We choose the ratio of horizontal and vertical emittances as $100:1$ i.e., $\bm{\varepsilon}_y$=$10^{-5}$ m-rad and $\bm{\varepsilon}_x$=$10^{-7}$ m-rad. This scheme requires an axial magnetic field $B$=71.87 Gauss at the position of the cathode, which can be generated by using either a permanent magnet or an electromagnet~\cite{KimPRSTB}. The current density at the cathode, which is required to produce an electron beam of desired emittances discussed above, is obtained by using the prescription given in Sec.~IV~A\textit{1} as $J_T$=0.12 A/$\text{cm}^2$ at $T$=2500 K. This value of current density is well within the range of a tungsten cathode~\cite{KimPRSTB}. Next, we discuss the requirements of the wiggler parameters to focus the flat beam described above. As given in Ref. \cite{Wigglerfield}, it is possible to generate a peak magnetic field upto 5 kG by using a hybrid wiggler with period 2.7 mm, and gap of 1 mm between the jaws of the wiggler. For a matched beam size $\sigma_x$=47 $\mu$m and vertical emittance $\bm{\varepsilon}_x$=$10^{-7}$ m-rad, we obtain $B_u$ = 1.1 kG. It has also been checked that the criteria given by Eq.~(\ref{eq:Extolerance}) is satisfied for this case.

Next, we discuss another possibility to relax the stringent requirement on the vertical beam emittance where, we can use a line cathode immersed in the solenoid field to produce a flat electron beam. This method has been discussed in detail in Sec.~IV~A\textit{2}. To produce a flat electron beam with beam half widths $\Delta x$=94 $\mu$m, and $\Delta y$=4.2 mm; we need a line cathode with $\Delta x_c$ = 94 $\mu$m, and $\Delta y_c$=4.2 mm. At $T$ = 2500 K, we obtain $\bm{\varepsilon}_x$ = $3.05\times 10^{-8}$ m-rad, and $\bm{\varepsilon}_y$ = $1.35\times 10^{-6}$ m-rad; which are quite acceptable. For beam current of 35 mA, the current density at the line cathode is obtained as $J_c$=2.86 A/$\text{cm}^2$, which can be easily achieved by the tungsten cathode for SEMs~\cite{SEMbook}. By using Eq.~(\ref{eq:Solenoidfield}), we find that the solenoid magnetic field $B(0)$ is required to be greater than 0.24 kG to focus such a flat electron beam. We choose $B(0)$= 1.0 kG. The Larmor radius is obtained as 10.15 $\mu$m for these parameters, which is significantly smaller than the rms beam size $\sigma_x$=47 $\mu$m. To keep the flat beam rotation less than 10 mrad over a length $L$=5 cm, we require the field uniformity $\Delta B/B$ to be better than 0.4 $\%$, as calculated by using Eq.~(\ref{eq:Rotationangle}).

%-------------------------------------------------------------------------------------------------------------------------------------------------------
\section{Discussions and conclusions}
\label{sec:discussion}
In this paper, we have presented a three-dimensional analysis of the surface mode in $\check{\text{C}}$erenkov FELs. Expressions have been derived for the electromagnetic field in a localized surface mode by suitably combining the plane wave solutions of Maxwell equations, propagating at different angles.
A crucial input for this calculation was to have the information about the change in $k_z$, after we include $e^{ik_yy}$ type dependence in the electromagnetic field, keeping the value of $\omega$ fixed. For the case of CFEL, this was simplified due to \textquotedblleft isotropic nature\textquotedblright~of the system in the $(y,z)$ plane, and in the case of SP-FEL, this was simplified due to the \textquotedblleft replacement rule\textquotedblright~for the evaluation of reflectivity of the incident evanescent wave. Interestingly, the \textquotedblleft isotropic nature\textquotedblright~is not applicable for SP-FEL case, and the \textquotedblleft replacement rule\textquotedblright~is not applicable for CFEL case. 

Three-dimensional analysis of the surface mode allows us to include the effect of diffraction, which plays an important role in the performance of a CFEL.
We have explained in the paper that for an isotropic system, as in the case of CFEL, in the diffraction term in the wave equation, $\lambda$ gets replaced with $\beta \lambda$. On the other hand if the system is not isotropic, but the EM field is present only in vacuum, as in the case of SP-FEL, $\lambda$ gets replaced with $\lambda/\beta_g$. Diffraction results in partial overlapping between the electron beam and the co-propagating optical mode. Consequently, the gain and the growth rate of a CFEL system reduce. To incorporate the 3D effects in the formulas for the gain and the growth rate calculations, we have taken the electron beam size to be same as the effective optical beam size, which has been evaluated by taking the 3D variations in the surface mode.

We also derived expressions for dielectric losses, and losses due to finite conductivity of the metal, which play an important role, when we increase the interaction length in order to increase the gain in a CFEL. Although all earlier analyses on single slab CFEL have ignored this effect, it is important to take such realistic effects into account in a practical device, as is the case in any device using guided waves at high frequency. It is interesting to point out that even in the case of SP-FELs, the effect of attenuation was neglected in earlier studies, and its importance was realized in later studies ~\cite{DispersionSPFEL,VinitFEL06}. In order to reduce the loss due to finite conductivity of metal in a CFEL, we have proposed that the metallic base can be kept at low temperature. We have optimized the parameters for a CFEL designed to operate at 0.1 THz, and shown that using a 46.5 keV electron beam with a current of 35 mA, an optimized CFEL oscillator can  deliver output power of 7.2 W at saturation with an efficiency of 0.44 $\%$. 

Our overall approach is built on the earlier analyses given in Refs.~\cite{KimPRSTB,VinitPRSTB}, where the diffraction properties of the surface mode have been studied to determine the requirements on electron beam parameters for the successful operation of an SP-FEL. Here, our aim has been to perform the analysis for CFEL case, and find out the electron beam parameters for a practical CFEL system. Like SP-FEL \cite{VinitPRSTB}, the requirements on the vertical beam emittance in a CFEL come out to be stringent, and we have discussed two ways to relax the stringent requirements. In the first scheme, a wiggler magnetic field is used to focus a flat electron beam, which is produced by a novel phase-space technique discussed in Ref.~\cite{KimPRSTB}. This scheme requires peak wiggler field of about 1.1 kG to focus a flat electron beam having transverse emittance ratio $\bm{\varepsilon}_y/\bm{\varepsilon}_x$=100. This value of transverse emittance ratio has been achieved recently at Fermi National Accelerator Laboratory \cite{FlatphaseE1}, and well below the recently proposed value of 400 \cite{Flatphase3}. In the second scheme, we used a solenoid field to focus the flat electron beam, which is produced by a line shaped tungsten cathode placed at the centre of solenoid. The solenoid field is taken as 1 kG with field uniformity $\Delta B/B$ required to be better than 0.4 $\%$ over a length of 5 cm. Although the suggested schemes may add to the complexity of the system, these are implementable, and are needed to satisfy the stringent requirements for optimum performance of the system.

We would like to mention that our analysis can be extended to understand the implications of 3D effects on working of a CFEL based on the other schemes such as - using double slab~\cite{EXPCFELD4,Li2} and negative refractive index material~\cite{Li3}. In the negative index material, the group velocity of the surface mode is negative. Due to the negative group velocity of the surface mode, CFEL system works like a BWO \cite{Li3}. $\check{\text{C}}$erenkov FEL in the BWO configuration can be studied by following an approach given in Refs.~\cite{VinitPRE,VinitPRSTB,KimPRSTB}, where the working of Smith-Purcell FELs in BWO configuration is discussed.

To summarize, we have performed 3D analysis of the surface mode and setup 3D Maxwell-Lorentz equations for a CFEL and SP-FEL system. We notice that the diffraction term appearing in Eqs.~(\ref{eq:MaxCFEL3D2}) and (\ref{eq:MaxSP3D}), is similar to the corresponding term for the case of undulator based FEL \cite{Braubook,Dattolibook}. Hence, the numerical techniques that are applied for solution of such equations in typical 3D FEL codes such as GENESIS~\cite{GENESIS} can be extended for 3D simulation of CFEL and SP-FEL, which may be taken up in future. The realistic effects such as the effect due to energy spread, and finite beam emittance can also be included in the computer code. We have optimized the parameters of a $\check{\text{C}}$erenkov FEL by including the 3D effects, and attenuation due to dielectric and ohmic losses; and find that the device can produce copious THz radiation, even after including these effects. Our analysis can be used for the detailed optimization of both CFEL and SP-FEL system.

%----------------------------------------------------------------------------------------------------------------------------------------------------------

\section*{Acknowledgment}
We acknowledge Professor S. B. Roy and Professor P. D. Gupta for constant encouragement. We also thank one unknown referee for very useful comments. One of us (YK) gratefully acknowledges Homi Bhabha National Institute, Department of Atomic Energy (India) for financial support during the work.

%----------------------------------------------------------------------------------------------------------------------------------------------------------
\appendix
\section{CALCULATIONS FOR THE ATTENUATION COEFFICIENT OF THE ELECTROMAGNETIC SURFACE MODE IN A CFEL}
\label{app:A}
In this Appendix, we calculate the attenuation coefficient of the electromagnetic surface mode due to loss inside the dielectric slab, and loss due to finite conductivity of the metal, which supports the dielectric slab. The schematic of the device is shown in Fig.~1, where free space is assigned as region I, dielectric slab as region II, and the metallic base as region III. The electromagnetic fields in region I are obtained by solving Maxwell equations with appropriate boundary conditions as~\cite{CFEL}:
\begin{eqnarray}
\label{eq:fieldvacuum}
H_y^I(x,z,t)=H\exp[i\psi-\Gamma(x+h)]+\text{c.c.}, ~~~~~~~~~~~~~~~~~~~\\
E_x^I(x,z,t)=(HZ_0/\beta)\exp[i\psi-\Gamma(x+h)]+\text{c.c.},~~~~~~~~~~\\
E_z^I(x,z,t)=(-i HZ_0/\beta\gamma)\exp[i\psi-\Gamma(x+h)]+\text{c.c.}.~~~~~~
\end{eqnarray}
Here, $H$ represents the magnetic field strength at the dielectric surface at $x$=$-h$. Electromagnetic fields in region II are given by~\cite{CFEL}:
\begin{eqnarray}
\label{eq:Hy2}
H_y^{II}(x,z,t)=\frac{ \epsilon\Gamma}{k_1}\frac{\cos[k_1(x+h+d)]}{\sin(k_1d)}H\exp(i\psi)+\text{c.c.},~~~~~
\end{eqnarray}
\begin{eqnarray}
\label{eq:Ex2}
E_x^{II}(x,z,t)=\frac{k_0\Gamma}{\omega \epsilon_0k_1}\frac{\cos[k_1(x+h+d)]}{\sin(k_1d)}H\exp(i\psi)+\text{c.c.},~~~~~
\end{eqnarray}
\begin{eqnarray}
\label{eq:Ez2}
E_z^{II}(x,z,t)=\frac{-i\Gamma}{\omega \epsilon_0}\frac{\sin[k_1(x+h+d)]}{\sin(k_1d)}H\exp(i\psi)+\text{c.c.},~~~~~~
\end{eqnarray}
where $k_1=\sqrt{\epsilon\omega^2/c^2-k_0^2}$. Total power transmitted by the electromagnetic fields is sum of power flow in region I and region II, which can be obtained by integrating the Poynting vector over the area transverse to the direction of beam propagation. The expression for total transmitted power is obtained as~\cite{CFEL}:
\begin{equation}
\label{eq:Pt}
\frac{P_t}{\Delta y}=\frac{Z_0H^2}{\beta k_0\epsilon^2a^2}[\gamma(1+\epsilon^2a^2)+\epsilon k_0d(1+a^2)].
\end{equation} 

Attenuation coefficient of the surface wave is given by \cite{Liao}: 
\begin{eqnarray}
\label{eq:attenuation1}
\alpha^{d,c}=\frac{P^{d,c}_l}{2P_t},
\end{eqnarray}
where, $P_l$ is power loss per unit length along the $z$-direction, and the superscripts $d$ and $c$ are meant for the dielectric and metallic conductor, respectively. In region II, namely the dielectric medium, losses are described with a complex relative permittivity $\tilde{\epsilon}=\epsilon'-i\epsilon''$, where $\tan\delta=\epsilon''/\epsilon'$ is identified as tangent loss of the dielectric medium~\cite{dielectricloss1}. Power loss per unit length in the dielectric medium can be written as~$P^{d}_l=\epsilon_0\epsilon\omega\tan\delta\int\big(\vert E^{II}_x\vert^2+\vert E^{II}_z\vert^2\big)dxdy$, where the integration is carried out from $-(h+d)$ to $-h$ in the $x$-direction and over length $\Delta y$ in the $y$-direction. By using Eqs.~(\ref{eq:Ex2}) and (\ref{eq:Ez2}), we obtain the expression for the power loss in the dielectric medium as:
\begin{eqnarray}
\label{eq:Pldielectric}
\frac{P^d_l}{\Delta y}=\frac{k_0H^2\tan\delta}{\epsilon_0\omega\epsilon^2a^2}[\gamma(2-\epsilon\beta^2)+\epsilon^2\beta^2k_0d(1+a^2)].
\end{eqnarray}
Now, Eqs.~(\ref{eq:Pt}) and (\ref{eq:Pldielectric}) are used in Eq.~(\ref{eq:attenuation1}) to obtain the dielectric attenuation coefficient as:
\begin{eqnarray}
\label{eq:alpha2}
\alpha^{d}=\frac{k_0\tan\delta}{2}\frac{[\gamma(2-\epsilon\beta^2)+\epsilon^2\beta^2k_0d(1+a^2)]}{[\gamma(1+\epsilon^2a^2)+\epsilon k_0d(1+a^2)]}.
\end{eqnarray}
In region III, which consist of metal, the dissipation of power occurs as ohmic losses due to finite conductivity of the metal.~Here, we have assumed that the dissipation occurs in a very small region near the metallic surface. The power loss per unit length along the metallic surface is given by $P_l^{c}$=$(R_s/2)\int \vert H_y\vert^2 dy$ \cite{Jackson}, where the integration is carried out over length $\Delta y$ in the $y$-direction. The magnitude of electromagnetic field $H_y$ at the metallic surface as given in Eq.~(\ref{eq:Hy2}) is used to evaluated $P_l^c$. By performing the required algebra for $P_l^c$, and using it along with Eq.~(\ref{eq:Pt}) for $P_t$ in Eq.~(\ref{eq:attenuation1}), the following expression for ohmic attenuation coefficient is obtained:
\begin{eqnarray}
\label{eq:alpha3}
\alpha^{c}=\frac{R_s}{Z_0}\frac{\beta\epsilon^2 k_0 (1+a^2)}{[\gamma(1+\epsilon^2a^2)+\epsilon k_0d(1+a^2)]}.
\end{eqnarray}
The sum of dielectric losses and losses due to finite conductivity of the metal gives total losses present in the system, and we write the total attenuation coefficient $\alpha$ of the surface mode as: $\alpha$=$\alpha^{d}+\alpha^{c}$.


\begin{thebibliography}{57}
\expandafter\ifx\csname natexlab\endcsname\relax\def\natexlab#1{#1}\fi
\expandafter\ifx\csname bibnamefont\endcsname\relax
  \def\bibnamefont#1{#1}\fi
\expandafter\ifx\csname bibfnamefont\endcsname\relax
  \def\bibfnamefont#1{#1}\fi
\expandafter\ifx\csname citenamefont\endcsname\relax
  \def\citenamefont#1{#1}\fi
\expandafter\ifx\csname url\endcsname\relax
  \def\url#1{\texttt{#1}}\fi
\expandafter\ifx\csname urlprefix\endcsname\relax\def\urlprefix{URL }\fi
\providecommand{\bibinfo}[2]{#2}
\providecommand{\eprint}[2][]{\url{#2}}

\bibitem[{\citenamefont{Danos et~al.}(1953)\citenamefont{Danos, Geschwind,
  Lashinsky, and Trier}}]{Danos1}
\bibinfo{author}{\bibfnamefont{M.}~\bibnamefont{Danos}},
  \bibinfo{author}{\bibfnamefont{S.}~\bibnamefont{Geschwind}},
  \bibinfo{author}{\bibfnamefont{H.}~\bibnamefont{Lashinsky}},
  \bibnamefont{and} \bibinfo{author}{\bibfnamefont{A.~V.} \bibnamefont{Trier}},
  \bibinfo{journal}{Phys. Rev.} \textbf{\bibinfo{volume}{92}},
  \bibinfo{pages}{828} (\bibinfo{year}{1953}).

\bibitem[{\citenamefont{Danos}(1955)}]{Danos}
\bibinfo{author}{\bibfnamefont{M.}~\bibnamefont{Danos}}, \bibinfo{journal}{J.
  Appl. Phys.} \textbf{\bibinfo{volume}{26}}, \bibinfo{pages}{2}
  (\bibinfo{year}{1955}).

\bibitem[{\citenamefont{Walsh et~al.}(1984)\citenamefont{Walsh, Johnson,
  Dattoli, and Renieri}}]{Walsh3}
\bibinfo{author}{\bibfnamefont{J.}~\bibnamefont{Walsh}},
  \bibinfo{author}{\bibfnamefont{B.}~\bibnamefont{Johnson}},
  \bibinfo{author}{\bibfnamefont{G.}~\bibnamefont{Dattoli}}, \bibnamefont{and}
  \bibinfo{author}{\bibfnamefont{A.}~\bibnamefont{Renieri}},
  \bibinfo{journal}{Phys. Rev. Lett.} \textbf{\bibinfo{volume}{53}},
  \bibinfo{pages}{779} (\bibinfo{year}{1984}).

\bibitem[{\citenamefont{Walsh et~al.}(1986)\citenamefont{Walsh, Johnson,
  Shaughnessy, Ciocci, Dattoli, Angelis, Dipace, Fiorentino, Gallerano, Letardi
  et~al.}}]{EXPCFEL0}
\bibinfo{author}{\bibfnamefont{J.~E.} \bibnamefont{Walsh}},
  \bibinfo{author}{\bibfnamefont{B.}~\bibnamefont{Johnson}},
  \bibinfo{author}{\bibfnamefont{C.}~\bibnamefont{Shaughnessy}},
  \bibinfo{author}{\bibfnamefont{F.}~\bibnamefont{Ciocci}},
  \bibinfo{author}{\bibfnamefont{G.}~\bibnamefont{Dattoli}},
  \bibinfo{author}{\bibfnamefont{A.}~\bibnamefont{Angelis}},
  \bibinfo{author}{\bibfnamefont{A.}~\bibnamefont{Dipace}},
  \bibinfo{author}{\bibfnamefont{E.}~\bibnamefont{Fiorentino}},
  \bibinfo{author}{\bibfnamefont{G.~P.} \bibnamefont{Gallerano}},
  \bibinfo{author}{\bibfnamefont{T.}~\bibnamefont{Letardi}},
  \bibnamefont{et~al.}, \bibinfo{journal}{Nucl. Instrum. Methods Phys. Res. A}
  \textbf{\bibinfo{volume}{250}}, \bibinfo{pages}{308} (\bibinfo{year}{1986}).

\bibitem[{\citenamefont{Ciocci et~al.}(1987)\citenamefont{Ciocci, Dattoli,
  Angelis, Dipace, Doria, Fiorentino, Gallerano, Letardi, Marino, Renieri
  et~al.}}]{EXPCFEL1}
\bibinfo{author}{\bibfnamefont{F.}~\bibnamefont{Ciocci}},
  \bibinfo{author}{\bibfnamefont{G.}~\bibnamefont{Dattoli}},
  \bibinfo{author}{\bibfnamefont{A.}~\bibnamefont{Angelis}},
  \bibinfo{author}{\bibfnamefont{A.}~\bibnamefont{Dipace}},
  \bibinfo{author}{\bibfnamefont{A.}~\bibnamefont{Doria}},
  \bibinfo{author}{\bibfnamefont{E.}~\bibnamefont{Fiorentino}},
  \bibinfo{author}{\bibfnamefont{G.~P.} \bibnamefont{Gallerano}},
  \bibinfo{author}{\bibfnamefont{T.}~\bibnamefont{Letardi}},
  \bibinfo{author}{\bibfnamefont{A.}~\bibnamefont{Marino}},
  \bibinfo{author}{\bibfnamefont{A.}~\bibnamefont{Renieri}},
  \bibnamefont{et~al.}, \bibinfo{journal}{Nucl. Instrum. Methods Phys. Res. A}
  \textbf{\bibinfo{volume}{259}}, \bibinfo{pages}{128} (\bibinfo{year}{1987}).

\bibitem[{\citenamefont{Garate et~al.}(1987)\citenamefont{Garate, Walsh,
  Shaughnessy, Johnson, and Moustaizis}}]{EXPCFELD1}
\bibinfo{author}{\bibfnamefont{E.~P.} \bibnamefont{Garate}},
  \bibinfo{author}{\bibfnamefont{J.}~\bibnamefont{Walsh}},
  \bibinfo{author}{\bibfnamefont{C.}~\bibnamefont{Shaughnessy}},
  \bibinfo{author}{\bibfnamefont{B.}~\bibnamefont{Johnson}}, \bibnamefont{and}
  \bibinfo{author}{\bibfnamefont{S.}~\bibnamefont{Moustaizis}},
  \bibinfo{journal}{Nucl. Instrum. Methods Phys. Res. A}
  \textbf{\bibinfo{volume}{259}}, \bibinfo{pages}{125} (\bibinfo{year}{1987}).

\bibitem[{\citenamefont{Walsh et~al.}(1988)\citenamefont{Walsh, Shaughnessy,
  Layman, Dattoli, Gallerano, and Renieri}}]{EXPCFELD2}
\bibinfo{author}{\bibfnamefont{J.}~\bibnamefont{Walsh}},
  \bibinfo{author}{\bibfnamefont{C.}~\bibnamefont{Shaughnessy}},
  \bibinfo{author}{\bibfnamefont{R.}~\bibnamefont{Layman}},
  \bibinfo{author}{\bibfnamefont{G.}~\bibnamefont{Dattoli}},
  \bibinfo{author}{\bibfnamefont{G.-P.} \bibnamefont{Gallerano}},
  \bibnamefont{and} \bibinfo{author}{\bibfnamefont{A.}~\bibnamefont{Renieri}},
  \bibinfo{journal}{Nucl. Instrum. Methods Phys. Res. A}
  \textbf{\bibinfo{volume}{272}}, \bibinfo{pages}{132} (\bibinfo{year}{1988}).

\bibitem[{\citenamefont{Ciocci et~al.}(1991)\citenamefont{Ciocci, Doria,
  Gallerano, Giabbai, Kimmitt, Messina, Renieri, and Walsh}}]{EXPCFEL2}
\bibinfo{author}{\bibfnamefont{F.}~\bibnamefont{Ciocci}},
  \bibinfo{author}{\bibfnamefont{A.}~\bibnamefont{Doria}},
  \bibinfo{author}{\bibfnamefont{G.~P.} \bibnamefont{Gallerano}},
  \bibinfo{author}{\bibfnamefont{I.}~\bibnamefont{Giabbai}},
  \bibinfo{author}{\bibfnamefont{M.~F.} \bibnamefont{Kimmitt}},
  \bibinfo{author}{\bibfnamefont{G.}~\bibnamefont{Messina}},
  \bibinfo{author}{\bibfnamefont{A.}~\bibnamefont{Renieri}}, \bibnamefont{and}
  \bibinfo{author}{\bibfnamefont{J.~E.} \bibnamefont{Walsh}},
  \bibinfo{journal}{Phys. Rev. Lett.} \textbf{\bibinfo{volume}{66}},
  \bibinfo{pages}{699} (\bibinfo{year}{1991}).

\bibitem[{\citenamefont{Fish and Walsh}(1992)}]{EXPCFEL3}
\bibinfo{author}{\bibfnamefont{E.~E.} \bibnamefont{Fish}} \bibnamefont{and}
  \bibinfo{author}{\bibfnamefont{J.~E.} \bibnamefont{Walsh}},
  \bibinfo{journal}{Appl. Phys. Lett.} \textbf{\bibinfo{volume}{60}},
  \bibinfo{pages}{1298} (\bibinfo{year}{1992}).

\bibitem[{\citenamefont{Seo. et~al.}(2000)\citenamefont{Seo., Choi, and
  Cho}}]{Seo}
\bibinfo{author}{\bibfnamefont{Y.}~\bibnamefont{Seo.}},
  \bibinfo{author}{\bibfnamefont{E.~H.} \bibnamefont{Choi}}, \bibnamefont{and}
  \bibinfo{author}{\bibfnamefont{G.~S.} \bibnamefont{Cho}},
  \bibinfo{journal}{J. Phys. D: Appl. Phys.} \textbf{\bibinfo{volume}{33}},
  \bibinfo{pages}{654} (\bibinfo{year}{2000}).

\bibitem[{\citenamefont{Owens and Brownell}(2003)}]{Owens1}
\bibinfo{author}{\bibfnamefont{I.~J.} \bibnamefont{Owens}} \bibnamefont{and}
  \bibinfo{author}{\bibfnamefont{J.~H.} \bibnamefont{Brownell}},
  \bibinfo{journal}{Phys. Rev. E} \textbf{\bibinfo{volume}{67}},
  \bibinfo{pages}{036611} (\bibinfo{year}{2003}).

\bibitem[{\citenamefont{Owens and Brownell}(2005)}]{Owens2}
\bibinfo{author}{\bibfnamefont{I.~J.} \bibnamefont{Owens}} \bibnamefont{and}
  \bibinfo{author}{\bibfnamefont{J.~H.} \bibnamefont{Brownell}},
  \bibinfo{journal}{J. Appl. Phys.} \textbf{\bibinfo{volume}{97}},
  \bibinfo{pages}{104915} (\bibinfo{year}{2005}).

\bibitem[{\citenamefont{Asakawa et~al.}(2006)\citenamefont{Asakawa, Nakao,
  Kusaba, and Tsunawaki}}]{EXPCFELD3}
\bibinfo{author}{\bibfnamefont{M.~R.} \bibnamefont{Asakawa}},
  \bibinfo{author}{\bibfnamefont{K.}~\bibnamefont{Nakao}},
  \bibinfo{author}{\bibfnamefont{M.}~\bibnamefont{Kusaba}}, \bibnamefont{and}
  \bibinfo{author}{\bibfnamefont{Y.}~\bibnamefont{Tsunawaki}}, in
  \emph{\bibinfo{booktitle}{Proceedings of FEL06}} (\bibinfo{year}{2006}), p.
  \bibinfo{pages}{364}.

\bibitem[{\citenamefont{Miyabe et~al.}(2007)\citenamefont{Miyabe, Ikeda,
  Asakawa, Kusaba, and Tsunawaki}}]{EXPCFELD4}
\bibinfo{author}{\bibfnamefont{N.}~\bibnamefont{Miyabe}},
  \bibinfo{author}{\bibfnamefont{A.}~\bibnamefont{Ikeda}},
  \bibinfo{author}{\bibfnamefont{M.~R.} \bibnamefont{Asakawa}},
  \bibinfo{author}{\bibfnamefont{M.}~\bibnamefont{Kusaba}}, \bibnamefont{and}
  \bibinfo{author}{\bibfnamefont{Y.}~\bibnamefont{Tsunawaki}}, in
  \emph{\bibinfo{booktitle}{Proceedings of FEL07}} (\bibinfo{year}{2007}), p.
  \bibinfo{pages}{406}.

\bibitem[{\citenamefont{Andrews and Brau}(2007)}]{Brau1}
\bibinfo{author}{\bibfnamefont{H.~L.} \bibnamefont{Andrews}} \bibnamefont{and}
  \bibinfo{author}{\bibfnamefont{C.~A.} \bibnamefont{Brau}},
  \bibinfo{journal}{J. Appl. Phys.} \textbf{\bibinfo{volume}{101}},
  \bibinfo{pages}{104904} (\bibinfo{year}{2007}).

\bibitem[{\citenamefont{Li et~al.}(2010)\citenamefont{Li, Huo, Imasaki,
  M.Asakawa, and Tsunawaki}}]{Li2}
\bibinfo{author}{\bibfnamefont{D.}~\bibnamefont{Li}},
  \bibinfo{author}{\bibfnamefont{G.}~\bibnamefont{Huo}},
  \bibinfo{author}{\bibfnamefont{K.}~\bibnamefont{Imasaki}},
  \bibinfo{author}{\bibnamefont{M.Asakawa}}, \bibnamefont{and}
  \bibinfo{author}{\bibfnamefont{Y.}~\bibnamefont{Tsunawaki}},
  \bibinfo{journal}{Infrared Phys. Techn.} \textbf{\bibinfo{volume}{53}},
  \bibinfo{pages}{204} (\bibinfo{year}{2010}).

\bibitem[{\citenamefont{Li et~al.}(2014)\citenamefont{Li, Wang, Hangyo, Wei,
  Yang, and Miyamoto}}]{Li3}
\bibinfo{author}{\bibfnamefont{D.}~\bibnamefont{Li}},
  \bibinfo{author}{\bibfnamefont{Y.}~\bibnamefont{Wang}},
  \bibinfo{author}{\bibfnamefont{M.}~\bibnamefont{Hangyo}},
  \bibinfo{author}{\bibfnamefont{Y.}~\bibnamefont{Wei}},
  \bibinfo{author}{\bibfnamefont{Z.}~\bibnamefont{Yang}}, \bibnamefont{and}
  \bibinfo{author}{\bibfnamefont{S.}~\bibnamefont{Miyamoto}},
  \bibinfo{journal}{Appl. Phys. Lett.} \textbf{\bibinfo{volume}{104}},
  \bibinfo{pages}{194102} (\bibinfo{year}{2014}).

\bibitem[{\citenamefont{Kalkal and Kumar}(2015)}]{CFEL}
\bibinfo{author}{\bibfnamefont{Y.}~\bibnamefont{Kalkal}} \bibnamefont{and}
  \bibinfo{author}{\bibfnamefont{V.}~\bibnamefont{Kumar}},
  \bibinfo{journal}{Phys. Rev. ST Accel. Beams} \textbf{\bibinfo{volume}{18}},
  \bibinfo{pages}{030707} (\bibinfo{year}{2015}).

\bibitem[{\citenamefont{Schachter and Ron}(1989)}]{Levi}
\bibinfo{author}{\bibfnamefont{L.}~\bibnamefont{Schachter}} \bibnamefont{and}
  \bibinfo{author}{\bibfnamefont{A.}~\bibnamefont{Ron}},
  \bibinfo{journal}{Phys. Rev. A} \textbf{\bibinfo{volume}{40}},
  \bibinfo{pages}{876} (\bibinfo{year}{1989}).

\bibitem[{\citenamefont{Hafizi et~al.}(1992)\citenamefont{Hafizi, Sprangle, and
  Serafim}}]{GratingFEL}
\bibinfo{author}{\bibfnamefont{B.}~\bibnamefont{Hafizi}},
  \bibinfo{author}{\bibfnamefont{P.}~\bibnamefont{Sprangle}}, \bibnamefont{and}
  \bibinfo{author}{\bibfnamefont{P.}~\bibnamefont{Serafim}},
  \bibinfo{journal}{Phys. Rev. A} \textbf{\bibinfo{volume}{45}},
  \bibinfo{pages}{8846} (\bibinfo{year}{1992}).

\bibitem[{\citenamefont{Urata et~al.}(1998)\citenamefont{Urata, Goldstein,
  Kimmitt, Naumov, Platt, and Walsh}}]{Urata}
\bibinfo{author}{\bibfnamefont{J.}~\bibnamefont{Urata}},
  \bibinfo{author}{\bibfnamefont{M.}~\bibnamefont{Goldstein}},
  \bibinfo{author}{\bibfnamefont{M.~F.} \bibnamefont{Kimmitt}},
  \bibinfo{author}{\bibfnamefont{A.}~\bibnamefont{Naumov}},
  \bibinfo{author}{\bibfnamefont{C.}~\bibnamefont{Platt}}, \bibnamefont{and}
  \bibinfo{author}{\bibfnamefont{J.~E.} \bibnamefont{Walsh}},
  \bibinfo{journal}{Phys. Rev. Lett.} \textbf{\bibinfo{volume}{80}},
  \bibinfo{pages}{516} (\bibinfo{year}{1998}).

\bibitem[{\citenamefont{Kim and Song}(2001)}]{Song}
\bibinfo{author}{\bibfnamefont{K.-J.} \bibnamefont{Kim}} \bibnamefont{and}
  \bibinfo{author}{\bibfnamefont{S.-B.} \bibnamefont{Song}},
  \bibinfo{journal}{Nucl. Instrum. Methods Phys. Res. A}
  \textbf{\bibinfo{volume}{475}}, \bibinfo{pages}{158} (\bibinfo{year}{2001}).

\bibitem[{\citenamefont{Andrews and Brau}(2004)}]{AB1}
\bibinfo{author}{\bibfnamefont{H.~L.} \bibnamefont{Andrews}} \bibnamefont{and}
  \bibinfo{author}{\bibfnamefont{C.~A.} \bibnamefont{Brau}},
  \bibinfo{journal}{Phys. Rev. ST Accel. Beams} \textbf{\bibinfo{volume}{7}},
  \bibinfo{pages}{070701} (\bibinfo{year}{2004}).

\bibitem[{\citenamefont{Andrews et~al.}(2005)\citenamefont{Andrews, Boulware,
  Brau, and Jarvis}}]{DispersionSPFEL}
\bibinfo{author}{\bibfnamefont{H.~L.} \bibnamefont{Andrews}},
  \bibinfo{author}{\bibfnamefont{C.~H.} \bibnamefont{Boulware}},
  \bibinfo{author}{\bibfnamefont{C.~A.} \bibnamefont{Brau}}, \bibnamefont{and}
  \bibinfo{author}{\bibfnamefont{J.~D.} \bibnamefont{Jarvis}},
  \bibinfo{journal}{Phys. Rev. ST Accel. Beams} \textbf{\bibinfo{volume}{8}},
  \bibinfo{pages}{050703} (\bibinfo{year}{2005}).

\bibitem[{\citenamefont{Kumar and Kim}(2006{\natexlab{a}})}]{VinitPRE}
\bibinfo{author}{\bibfnamefont{V.}~\bibnamefont{Kumar}} \bibnamefont{and}
  \bibinfo{author}{\bibfnamefont{K.-J.} \bibnamefont{Kim}},
  \bibinfo{journal}{Phys. Rev. E} \textbf{\bibinfo{volume}{73}},
  \bibinfo{pages}{026501} (\bibinfo{year}{2006}{\natexlab{a}}).

\bibitem[{\citenamefont{Kim and Kumar}(2007)}]{KimPRSTB}
\bibinfo{author}{\bibfnamefont{K.-J.} \bibnamefont{Kim}} \bibnamefont{and}
  \bibinfo{author}{\bibfnamefont{V.}~\bibnamefont{Kumar}},
  \bibinfo{journal}{Phys. Rev. ST Accel. Beams} \textbf{\bibinfo{volume}{10}},
  \bibinfo{pages}{080702} (\bibinfo{year}{2007}).

\bibitem[{\citenamefont{Kumar and Kim}(2009)}]{VinitPRSTB}
\bibinfo{author}{\bibfnamefont{V.}~\bibnamefont{Kumar}} \bibnamefont{and}
  \bibinfo{author}{\bibfnamefont{K.-J.} \bibnamefont{Kim}},
  \bibinfo{journal}{Phys. Rev. ST Accel. Beams} \textbf{\bibinfo{volume}{12}},
  \bibinfo{pages}{070703} (\bibinfo{year}{2009}).

\bibitem[{\citenamefont{Kumar and Kim}(2007)}]{VinitFEL07}
\bibinfo{author}{\bibfnamefont{V.}~\bibnamefont{Kumar}} \bibnamefont{and}
  \bibinfo{author}{\bibfnamefont{K.-J.} \bibnamefont{Kim}}, in
  \emph{\bibinfo{booktitle}{Proceedings of FEL07}} (\bibinfo{year}{2007}), pp.
  \bibinfo{pages}{38--41}.

\bibitem[{\citenamefont{Gardelle et~al.}(2009)\citenamefont{Gardelle, Courtois,
  Modin, and Donohue}}]{EXPSPFEL1}
\bibinfo{author}{\bibfnamefont{J.}~\bibnamefont{Gardelle}},
  \bibinfo{author}{\bibfnamefont{L.}~\bibnamefont{Courtois}},
  \bibinfo{author}{\bibfnamefont{P.}~\bibnamefont{Modin}}, \bibnamefont{and}
  \bibinfo{author}{\bibfnamefont{J.~T.} \bibnamefont{Donohue}},
  \bibinfo{journal}{Phys. Rev. ST Accel. Beams} \textbf{\bibinfo{volume}{12}},
  \bibinfo{pages}{110701} (\bibinfo{year}{2009}).

\bibitem[{\citenamefont{Andrews et~al.}(2009)\citenamefont{Andrews, Brau,
  Jarvis, Guertin, O'Donnell, Durant, Lowell, and Mross}}]{EXPSPFEL2}
\bibinfo{author}{\bibfnamefont{H.~L.} \bibnamefont{Andrews}},
  \bibinfo{author}{\bibfnamefont{C.~A.} \bibnamefont{Brau}},
  \bibinfo{author}{\bibfnamefont{J.~D.} \bibnamefont{Jarvis}},
  \bibinfo{author}{\bibfnamefont{C.~F.} \bibnamefont{Guertin}},
  \bibinfo{author}{\bibfnamefont{A.}~\bibnamefont{O'Donnell}},
  \bibinfo{author}{\bibfnamefont{B.}~\bibnamefont{Durant}},
  \bibinfo{author}{\bibfnamefont{T.~H.} \bibnamefont{Lowell}},
  \bibnamefont{and} \bibinfo{author}{\bibfnamefont{M.~R.} \bibnamefont{Mross}},
  \bibinfo{journal}{Phys. Rev. ST Accel. Beams} \textbf{\bibinfo{volume}{12}},
  \bibinfo{pages}{080703} (\bibinfo{year}{2009}).

\bibitem[{\citenamefont{Donohue and Gardelle}(2011)}]{Donohue3DSPFEL}
\bibinfo{author}{\bibfnamefont{J.~T.} \bibnamefont{Donohue}} \bibnamefont{and}
  \bibinfo{author}{\bibfnamefont{J.}~\bibnamefont{Gardelle}},
  \bibinfo{journal}{Phys. Rev. ST Accel. Beams} \textbf{\bibinfo{volume}{14}},
  \bibinfo{pages}{060709} (\bibinfo{year}{2011}).

\bibitem[{\citenamefont{Clery}(2002)}]{Clery}
\bibinfo{author}{\bibfnamefont{D.}~\bibnamefont{Clery}},
  \bibinfo{journal}{Science} \textbf{\bibinfo{volume}{297}},
  \bibinfo{pages}{761} (\bibinfo{year}{2002}).

\bibitem[{\citenamefont{Siegel}(2002)}]{Siegel}
\bibinfo{author}{\bibfnamefont{P.~H.} \bibnamefont{Siegel}},
  \bibinfo{journal}{IEEE Trans. Microwave Theory Tech.}
  \textbf{\bibinfo{volume}{50}}, \bibinfo{pages}{910} (\bibinfo{year}{2002}).

\bibitem[{\citenamefont{Siegman}(1986)}]{Lasers}
\bibinfo{author}{\bibfnamefont{A.~E.} \bibnamefont{Siegman}},
  \emph{\bibinfo{title}{Lasers}} (\bibinfo{publisher}{University Science books
  Mil Valley, California}, \bibinfo{year}{1986}).

\bibitem[{\citenamefont{Kim}(1989)}]{Kimsynchroton}
\bibinfo{author}{\bibfnamefont{K.-J.} \bibnamefont{Kim}}, in
  \emph{\bibinfo{booktitle}{AIP Conference Proceedings No. 184 (AIP, New
  York)}} (\bibinfo{year}{1989}), pp. \bibinfo{pages}{565--632}.

\bibitem[{\citenamefont{Pampaloni and Enderlein}(2004)}]{Primer}
\bibinfo{author}{\bibfnamefont{F.}~\bibnamefont{Pampaloni}} \bibnamefont{and}
  \bibinfo{author}{\bibfnamefont{J.}~\bibnamefont{Enderlein}}
  (\bibinfo{year}{2004}), \eprint{arXiv:physics/0410021}.

\bibitem[{\citenamefont{Horikis and McCallum}(2006)}]{Selffourier}
\bibinfo{author}{\bibfnamefont{T.~P.} \bibnamefont{Horikis}} \bibnamefont{and}
  \bibinfo{author}{\bibfnamefont{M.~S.} \bibnamefont{McCallum}},
  \bibinfo{journal}{J. Opt. Soc. Am. A} \textbf{\bibinfo{volume}{23}},
  \bibinfo{pages}{829} (\bibinfo{year}{2006}).

\bibitem[{\citenamefont{Kumar and Kim}(2006{\natexlab{b}})}]{VinitFEL06}
\bibinfo{author}{\bibfnamefont{V.}~\bibnamefont{Kumar}} \bibnamefont{and}
  \bibinfo{author}{\bibfnamefont{K.-J.} \bibnamefont{Kim}}, in
  \emph{\bibinfo{booktitle}{Proceedings of FEL06}}
  (\bibinfo{year}{2006}{\natexlab{b}}), pp. \bibinfo{pages}{67--70}.

\bibitem[{\citenamefont{Kumar and Kim}(2005)}]{VinitPAC05}
\bibinfo{author}{\bibfnamefont{V.}~\bibnamefont{Kumar}} \bibnamefont{and}
  \bibinfo{author}{\bibfnamefont{K.-J.} \bibnamefont{Kim}}, in
  \emph{\bibinfo{booktitle}{Proceedings of the 21st Particle Accelerator
  Conference, Knoxville, TN}} (\bibinfo{publisher}{IEEE, Piscataway, NJ},
  \bibinfo{year}{2005}), pp. \bibinfo{pages}{1616--1618}.

\bibitem[{\citenamefont{Brau}(1990)}]{Braubook}
\bibinfo{author}{\bibfnamefont{C.~A.} \bibnamefont{Brau}},
  \emph{\bibinfo{title}{Free-electron laser}} (\bibinfo{publisher}{Academic
  Press, San Diego}, \bibinfo{year}{1990}).

\bibitem[{\citenamefont{Chao}(1993)}]{Chaobook}
\bibinfo{author}{\bibfnamefont{A.}~\bibnamefont{Chao}},
  \emph{\bibinfo{title}{Physics of Collective Beam Instabilities in High Energy
  Accelerators}} (\bibinfo{publisher}{John Wiley $\&$ Sons, Inc., New York},
  \bibinfo{year}{1993}).

\bibitem[{\citenamefont{Piot et~al.}(2006)\citenamefont{Piot, Sun, and
  Kim}}]{FlatphaseE1}
\bibinfo{author}{\bibfnamefont{P.}~\bibnamefont{Piot}},
  \bibinfo{author}{\bibfnamefont{Y.-E.} \bibnamefont{Sun}}, \bibnamefont{and}
  \bibinfo{author}{\bibfnamefont{K.-J.} \bibnamefont{Kim}},
  \bibinfo{journal}{Phys. Rev. ST Accel. Beams} \textbf{\bibinfo{volume}{9}},
  \bibinfo{pages}{031001} (\bibinfo{year}{2006}).

\bibitem[{\citenamefont{Scharlemann}(1985)}]{Flatfocusing1}
\bibinfo{author}{\bibfnamefont{E.}~\bibnamefont{Scharlemann}},
  \bibinfo{journal}{J. Appl. Phys.} \textbf{\bibinfo{volume}{58}},
  \bibinfo{pages}{2154} (\bibinfo{year}{1985}).

\bibitem[{\citenamefont{Booske et~al.}(1988)\citenamefont{Booske, Destler,
  Segalov, Radack, Rosenbury, Rodgers, Jr., Granatstein, and
  Mayergoyz}}]{Flatfocusing2}
\bibinfo{author}{\bibfnamefont{J.~H.} \bibnamefont{Booske}},
  \bibinfo{author}{\bibfnamefont{W.~W.} \bibnamefont{Destler}},
  \bibinfo{author}{\bibfnamefont{Z.}~\bibnamefont{Segalov}},
  \bibinfo{author}{\bibfnamefont{D.~J.} \bibnamefont{Radack}},
  \bibinfo{author}{\bibfnamefont{E.~T.} \bibnamefont{Rosenbury}},
  \bibinfo{author}{\bibfnamefont{J.}~\bibnamefont{Rodgers}},
  \bibinfo{author}{\bibfnamefont{T.~M.~A.} \bibnamefont{Jr.}},
  \bibinfo{author}{\bibfnamefont{V.~L.} \bibnamefont{Granatstein}},
  \bibnamefont{and} \bibinfo{author}{\bibfnamefont{I.~D.}
  \bibnamefont{Mayergoyz}}, \bibinfo{journal}{J. Appl. Phys.}
  \textbf{\bibinfo{volume}{64}}, \bibinfo{pages}{6} (\bibinfo{year}{1988}).

\bibitem[{\citenamefont{Li et~al.}(2006)\citenamefont{Li, Yang, Imasaki, and
  Park}}]{LiSolenoid}
\bibinfo{author}{\bibfnamefont{D.}~\bibnamefont{Li}},
  \bibinfo{author}{\bibfnamefont{Z.}~\bibnamefont{Yang}},
  \bibinfo{author}{\bibfnamefont{K.}~\bibnamefont{Imasaki}}, \bibnamefont{and}
  \bibinfo{author}{\bibfnamefont{G.-S.} \bibnamefont{Park}},
  \bibinfo{journal}{Phys. Rev. ST Accel. Beams} \textbf{\bibinfo{volume}{9}},
  \bibinfo{pages}{040701} (\bibinfo{year}{2006}).

\bibitem[{\citenamefont{Donohue and Gardelle}(2006)}]{DonohueSolenoid}
\bibinfo{author}{\bibfnamefont{J.~T.} \bibnamefont{Donohue}} \bibnamefont{and}
  \bibinfo{author}{\bibfnamefont{J.}~\bibnamefont{Gardelle}},
  \bibinfo{journal}{Phys. Rev. ST Accel. Beams} \textbf{\bibinfo{volume}{9}},
  \bibinfo{pages}{060701} (\bibinfo{year}{2006}).

\bibitem[{\citenamefont{Brinkmann et~al.}(2001)\citenamefont{Brinkmann,
  Derbenev, and Fl\"{o}ttmann}}]{Flatproduction}
\bibinfo{author}{\bibfnamefont{R.}~\bibnamefont{Brinkmann}},
  \bibinfo{author}{\bibfnamefont{Y.}~\bibnamefont{Derbenev}}, \bibnamefont{and}
  \bibinfo{author}{\bibfnamefont{K.}~\bibnamefont{Fl\"{o}ttmann}},
  \bibinfo{journal}{Phys. Rev. ST Accel. Beams} \textbf{\bibinfo{volume}{4}},
  \bibinfo{pages}{053501} (\bibinfo{year}{2001}).

\bibitem[{\citenamefont{Westphal and Sils}(1972)}]{dielectriclossdata}
\bibinfo{author}{\bibfnamefont{W.}~\bibnamefont{Westphal}} \bibnamefont{and}
  \bibinfo{author}{\bibfnamefont{A.}~\bibnamefont{Sils}}, \bibinfo{type}{Tech.
  Rep.} \bibinfo{number}{AFML-TR-72-39}, \bibinfo{institution}{Massachusetts
  Institute of Technology} (\bibinfo{year}{1972}).

\bibitem[{\citenamefont{Tanner and Larson}(1968)}]{Silverresitivity}
\bibinfo{author}{\bibfnamefont{D.~B.} \bibnamefont{Tanner}} \bibnamefont{and}
  \bibinfo{author}{\bibfnamefont{D.~C.} \bibnamefont{Larson}},
  \bibinfo{journal}{Phys. Rev.} \textbf{\bibinfo{volume}{166}},
  \bibinfo{pages}{652} (\bibinfo{year}{1968}).

\bibitem[{\citenamefont{Zhu et~al.}(2014)\citenamefont{Zhu, Piot, Mihalcea, and
  Prokop}}]{Flatphase3}
\bibinfo{author}{\bibfnamefont{J.}~\bibnamefont{Zhu}},
  \bibinfo{author}{\bibfnamefont{P.}~\bibnamefont{Piot}},
  \bibinfo{author}{\bibfnamefont{D.}~\bibnamefont{Mihalcea}}, \bibnamefont{and}
  \bibinfo{author}{\bibfnamefont{C.~R.} \bibnamefont{Prokop}},
  \bibinfo{journal}{Phys. Rev. ST Accel. Beams} \textbf{\bibinfo{volume}{17}},
  \bibinfo{pages}{084401} (\bibinfo{year}{2014}).

\bibitem[{\citenamefont{Papadichev and Rybalchenko}(1998)}]{Wigglerfield}
\bibinfo{author}{\bibfnamefont{V.~A.} \bibnamefont{Papadichev}}
  \bibnamefont{and} \bibinfo{author}{\bibfnamefont{G.~V.}
  \bibnamefont{Rybalchenko}}, \bibinfo{journal}{Nucl. Instrum. Methods Phys.
  Res. A} \textbf{\bibinfo{volume}{407}}, \bibinfo{pages}{419}
  (\bibinfo{year}{1998}).

\bibitem[{\citenamefont{Reimer and Kohl}(1999)}]{SEMbook}
\bibinfo{author}{\bibfnamefont{L.}~\bibnamefont{Reimer}} \bibnamefont{and}
  \bibinfo{author}{\bibfnamefont{H.}~\bibnamefont{Kohl}},
  \emph{\bibinfo{title}{Transmission Electron Microscopy}}
  (\bibinfo{publisher}{Springer Series in Optical Sciences (Springer Verlag,
  Berlin), Vol. 36}, \bibinfo{year}{1999}).

\bibitem[{\citenamefont{Dattoli et~al.}(1993)\citenamefont{Dattoli, Renieri,
  and Torre}}]{Dattolibook}
\bibinfo{author}{\bibfnamefont{G.}~\bibnamefont{Dattoli}},
  \bibinfo{author}{\bibfnamefont{A.}~\bibnamefont{Renieri}}, \bibnamefont{and}
  \bibinfo{author}{\bibfnamefont{A.}~\bibnamefont{Torre}},
  \emph{\bibinfo{title}{Lectrures on the Free Electron Laser Theory and related
  Topics}} (\bibinfo{publisher}{World Scientific, Singapore},
  \bibinfo{year}{1993}).

\bibitem[{\citenamefont{Reiche}(1999)}]{GENESIS}
\bibinfo{author}{\bibfnamefont{S.}~\bibnamefont{Reiche}},
  \bibinfo{journal}{Nucl. Instrum. Methods Phys. Res. A}
  \textbf{\bibinfo{volume}{429}}, \bibinfo{pages}{243} (\bibinfo{year}{1999}).

\bibitem[{\citenamefont{Liao}(2011)}]{Liao}
\bibinfo{author}{\bibfnamefont{S.~Y.} \bibnamefont{Liao}},
  \emph{\bibinfo{title}{Microwave Devices and Circuits}}
  (\bibinfo{publisher}{Dorling Kindersley, India}, \bibinfo{year}{2011}).

\bibitem[{\citenamefont{Waldron}(1970)}]{dielectricloss1}
\bibinfo{author}{\bibfnamefont{R.~A.} \bibnamefont{Waldron}},
  \emph{\bibinfo{title}{Theory of guided electromagnetic waves}}
  (\bibinfo{publisher}{London ; New York : Van Nostrand Reinhold},
  \bibinfo{year}{1970}).

\bibitem[{\citenamefont{Jackson}(1999)}]{Jackson}
\bibinfo{author}{\bibfnamefont{J.~D.} \bibnamefont{Jackson}},
  \emph{\bibinfo{title}{Classical Electrodynamics}} (\bibinfo{publisher}{John
  Wiley, Singapore}, \bibinfo{year}{1999}).

\end{thebibliography}
\end{document}